\documentclass[twocolumn,preprintnumbers]{revtex4}

\usepackage{graphicx}
\usepackage{epsf}
\usepackage{amsmath,amssymb}
\newcommand{\bvec}{\boldsymbol}

\begin{document}
\preprint{KUNS-2503}

\title{Parity mixing of pair at nuclear surface due to spin-orbit potential
in $^{18}$F}
\author{Yoshiko Kanada-En'yo}
\affiliation{Department of Physics, Kyoto University, Kyoto 606-8502, Japan}
\author{Fumiharu Kobayashi}
\affiliation{Department of Physics, Niigata University, Niigata 950-2181, Japan}
\begin{abstract}
We investigate the structure of $^{18}$F with the microscopic wave function based
on the three-body $^{16}$O+$p$+$n$ model.
In the calculation of the generator coordinate method (GCM) of the three-body model, $T=0$ energy spectra of 
$J^\pi=1^+$, $3^+$, and $5^+$ states and $T=1$ spectra of 
$J^\pi=0^+$, $2^+$ states in $^{18}$F are described reasonably. 
Based on the dinucleon picture, 
the effect of the spin-orbit force 
on the $T=0$ and $T=1$ $pn$ pairs around the $^{16}$O core is discussed.
The $T=1$ pair in the $J^\pi=0^+$ state gains the spin-orbit potential energy 
involving the odd-parity mixing in the pair. 
The spin-orbit potential energy gain with the parity mixing is not so efficient 
for the $T=0$ pair in the $J^\pi=1^+$ state.  
The parity mixing in the pair is regarded
as the internal symmetry breaking of the pair in the 
spin-orbit potential at the nuclear surface.
\end{abstract}

\maketitle
\section{Introduction}

Motivated by the progress of experimental researches on 
unstable nuclei, the interest on 
proton and neutron ($pn$) pair correlations has been revived in these years, 
and $pn$
pairing in proton-rich nuclei near the $Z=N$ line has been intensively investigated. 
The importance of the $pn$ pairing has been discussed 
for a long time (see Refs.~\cite{Goodman79,Goodman:1999zz,VanIsacker:2013rva} and references therein).
In the recent studies of the $pn$ pairing, the competition between isosalar $T=0$ pairing 
and isovector $T=1$ pairing has been attracting a great interest. 
Moreover, the possibility of the $T=0$ and $T=1$ mixed pairing  
has been studied with the recently developed mean-field approaches
including the isospin mixing\cite{Satula:2009qu,Gezerlis:2011rh,Sato:2013yca}. 
The $pn$ pairing has been also discussed 
in relation with the Wigner energy in even-even nuclei, and more explicitly, 
its role in four-nucleon $\alpha$-like correlations has been investigated
in recent works \cite{Sandulescu:2012xv,Sandulescu:2012jm}.

The competition between $T=0$ pairing 
and $T=1$ pairing is one of the essential problems in $Z\sim N$ nuclei. 
As well known, the nuclear interaction in free space is 
more attractive in the $T=0$ spin-triplet 
even ($^3$E) channel than in the $T=1$ spin-singlet even ($^1$E) channel, 
as known in the fact that the $T=0$ $pn$ system forms a bound state, deuteron. 
Since the $T=0$ interaction is considered to be stronger than 
the $T=1$ interaction also in nuclear medium and at the nuclear surface, 
it is expected naively that the deuteron-like $T=0$ pair is more favored 
than the $T=1$ pair. 
Nevertheless, as seen in the ground state spins of the 
odd-odd nuclei, the $T=1$ pairing is often favored than the $T=0$ pairing except for light nuclei
\cite{Macchiavelli:1999kf}.
Many theoretical calculations have been achieved 
to investigate the competition between the $T=0$ and $T=1$ pairing and suggested
that the $T=1$ pairing rather than the $T=0$ pairing occurs in the ground states of many $Z=N$ medium-mass nuclei 
\cite {Gezerlis:2011rh,Engel:1996sh,Satula:1996dc,Poves:1998br,Goodman:1998zz,Kaneko:2004ya,Baroni:2009eh,Bertsch:2009xz,Sagawa:2012ta}.
The origin of the suppression of the $T=0$ pairing has been discussed 
from the standpoint of the coupling of single-particle angular momenta, 
and the important role of the spin-orbit force in the $T=0$ and $T=1$ pair 
competition has been pointed out \cite{Poves:1998br,Baroni:2009eh,Bertsch:2009xz,Gezerlis:2011rh,Sagawa:2012ta,Bertsch:2009zd}.  
Namely, the spin-orbit potential tends to suppress more 
the $T=0$ $J=1$ pairing than the $T=1$ $J=0$ pairing because 
the $S$-wave component of the relative motion between two nucleons in the 
$j^2$ configuration is much smaller for the $T=0$ $J=1$ pair 
 than for the $T=1$ $J=0$ pair. 

On the other hand, it has been theoretically suggested 
that, in rotational states, the $T=0$ pairing is more favored than the $T=1$ pairing 
in the high spin region 
\cite{Nichols78,Muhlhans81,Kaneko:1998zz,Terasaki:1998bp,Goodman:2001gx,Hasegawa:2004br}.
Recently, a spin-aligned $pn$ pair in 
medium-mass $N=Z$ nuclei has been reported in the experimental work
\cite{Cederwall:2011dt}
which has stimulated the subsequent theoretical studies \cite{Zerguine:2011zb,Qi:2011yj}.

The trend of the $pn$ pairing condensation can be understood 
by the feature of one $pn$ pair. That is, a $T=0$ $pn$ pair is 
suppressed in low spin states while it can persist in a rotating system.
In the recent work by Tanimura {\it et al.} \cite{Tanimura:2013cea}, 
the $pn$ correlation in a single pair around a core has been systematically 
studied within the three-body potential model calculation, and it has been shown that
 the spin-orbit potential plays an important role in  
the suppression of the $pn$ pair in the $T=0$ $J^\pi=1^+$ state.
It means that the properties of a $pn$ pair around a core reflects the basic feature of the 
$pn$ correlation, and the study of one $pn$ pair can be helpful to understand the 
$pn$ pairing phenomena in nuclei.

Our aim is to investigate features of a $pn$ pair around a core nucleus
and clarify the mechanism how the spin-orbit field affects the 
$pn$ pair. In the present work, we discuss the effect of the spin-orbit force
from the standpoint of the two-nucleon cluster (dinucleon) picture.
Let us consider a proton and a neutron in the potential given by 
the core nucleus. 
It is expected that the proton and the neutron form a deuteron-like $T=0$ pair 
at the nuclear surface because of the nucleon-nucleon attraction 
in the $^3$E channel.  If there is no spin-orbit potential, 
the $T=0$ pair should be the pure spin-triplet $S=1$ state. Similarly to the 
$(TS)=(01)$ pair, because of the $^1$E attraction
a proton and a neutron in the $T=1$ state may form 
a $(TS)=(10)$ pair, which is the analog state of the $nn$ pair in a system of
two neutrons around a core. 
The $T=0$ and $T=1$ $pn$ pairs are not necessarily same as the two-nucleon (quasi)bound states in 
free space, but the pairs are regarded as virtual 
bound states of spatially correlating two nucleons 
confined in the central potential from the core. 
We regard such a pair as a dinucleon cluster.

In the central potential without the spin-orbit potential,
the Hamiltonian for two nucleons is invariant with respect to the internal parity 
transformation of the pair, which is equivalent to the exchange 
of the coordinates of the first and the second nucleons $\bvec{r}_1\leftrightarrow\bvec{r}_2$ in the pair. The lowest $T=0$ and $T=1$ states 
of two nucleons are the $(TS)=(01)$ and $(TS)=(10)$ states which are pure 
even-parity states without the odd-parity mixing because of the symmetry of the Hamiltonian. 
Then, considering the spin-orbit potential as a perturbative field, 
we discuss how the $T=0$ and $T=1$ $pn$ pairs are affected by 
the external perturbative field.
This picture is different from the usual mean-field picture in the $jj$ coupling scheme, 
in which single-particle orbits in 
the central and spin-orbit potentials are considered and 
pair correlations are caused by the residual $pn$ interaction (see appendix \ref{sec:app1}).

\begin{figure}[tb]
\begin{center}
\includegraphics[width=6.5cm]{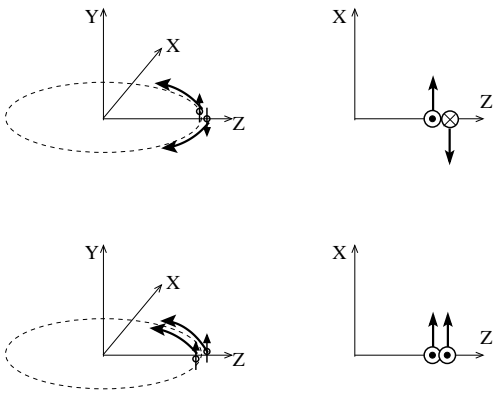} 	
\end{center}
\vspace{0.5cm}
  \caption{Schematic figures for two nucleons in the pair in the spin-orbit potential
  at the nuclear surface. The $\uparrow_Y\downarrow_Y$ $pn$ pair and 
  the $\uparrow_Y\uparrow_Y$ $pn$ pair in the body-fixed frame $XYZ$ are 
  shown in the upper and lower panels, respectively.  
\label{fig:pn-pair}}
\end{figure}

The present work is based on the following idea in 
the dinucleon picture.
Suppose that two nucleons in the $(TS)=(01)$ pair are localized 
around a position on the $Z$ axis and they have parallel intrinsic spins 
oriented to the $Y$ axis (see lower panels of Fig.~\ref{fig:pn-pair}). 
When the perturbative field of 
the spin-orbit potential is imposed, two nucleons in the pair are boosted to have 
finite momenta along the $X$ axis, i.e., the finite orbital angular momenta 
around the core so as to gain the spin-orbit potential energy. 
The boosting of two nucleons in the same direction 
does not cause the internal structure change of the pair but it changes only 
the center of mass (c.m.) motion of the pair. In the total system,  
the c.m. motion of the pair along the $X$ axis corresponds to the 
rotational mode of the pair around the core. 
It means that high spin states gain the spin-orbit potential energy without
the internal energy loss of the pair. 
On the other hand, the low spin state
with zero orbital angular momentum $L=0$ can not gain the spin-orbit potential. 
Thus the dinucleon picture provides a simple interpretation 
for the feature that the $T=0$ $pn$ pair
can persist in a rotating system but it
may be relatively unfavored by the spin-orbit force in the low spin state.

Moreover, based on the dinucleon picture, 
we find further interesting phenomenon concerning the symmetry breaking 
of the $T=1$ $pn$ pair, that is, the parity mixing in the $T=1$ pair. 
Suppose that two nucleons in the $(TS)=(10)$ pair are localized 
around a position on the $Z$ axis and they have antiparallel intrinsic spins 
oriented to the $Y$ axis (see upper panels of Fig.~\ref{fig:pn-pair}). 
When 
the spin-orbit potential is imposed, 
the internal parity symmetry is explicitly broken by the perturbative field, 
because the spin-orbit potential is not invariant with respect to the transformation 
$\bvec{r}_1\leftrightarrow\bvec{r}_2$ for the spin-up and -down nucleons
in the pair. 
Actually, in the potential with the spin-orbit field, 
spin-up and -down nucleons in the pair 
are boosted along the $X$ axis in the opposite direction
so as to gain the spin-orbit potential energy. As a result of the boosting in the 
opposite direction, the internal structure of the pair changes and 
the odd-parity component mixes into the dominant even-parity component
in the internal pair wave function.
Namely, the parity mixing of $(TS)=(10)$ and $(11)$ occurs in the $T=1$ pair.
Since the odd-parity mixing causes the internal energy loss of the pair, 
the odd-parity component is determined by the competition 
between the spin-orbit potential energy gain and the internal energy loss. 
Strictly speaking, the $pn$ pair of antiparallel spin nucleons contains 
the $T=0$ component as well as the $T=1$ component, and in principle, 
the parity mixing of $(TS)=(01)$ and $(00)$ may also occur in the $T=0$ pair 
because of the spin-orbit potential energy gain. 
However, as shown in this paper, 
the spin-orbit potential energy gain 
with the parity mixing is suppressed in the $T=0$ channel because 
the $T=0$ pair tends to lose much 
internal energy than the $T=1$ pair even in the case of the same $^3$E and $^1$E attractions.

Similar phenomenon of parity mixing of pairs has been discussed
recently in the condensed matter physics \cite{Fujimoto09}.
The parity mixing of Cooper pairs has been suggested 
to occur because of the spin-orbit interaction 
in noncentrosymmetric superconductors having the breaking of 
inversion symmetry in the crystal structure.
The mechanism of the parity mixing is analogous to that of the two-nucleon pair 
caused by the spin-orbit field at the nuclear surface. 

In this paper, we investigate the structure of $^{18}$F 
with the microscopic wave function based on a
three-body $^{16}{\rm O}+p+n$ model
and discuss the behavior of the $pn$ pair around $^{16}$O. 
In the present model, two nucleons are treated as 
Gaussian wave packets around $^{16}{\rm O}$ which is 
assumed to be the inert core written by 
the harmonic oscillator (H.O.) $p$-shell closed configuration.
The antisymmetrization between 18 nucleons and the recoil effect of the 
core are exactly taken into account, and 
the energy spectra of $^{18}$F are calculated using phenomenological 
effective nuclear forces. 
In that sense, the $^{18}$F wave function used in the present work is 
fully microscopic one. 
One of the advantages of the present model is that 
the expression of the Gaussian wave packets for 
valence nucleon wave functions provides the direct link with
the $(0s)^2$ $pn$ cluster formation and its breaking at the nuclear 
surface. Moreover, the internal wave function and the c.m. motion of the pair
are separable.  It is also helpful to consider a classical picture for 
the position and momentum of valence nucleons in the pair.
Based on the dinucleon picture, we analyze the effect of the spin-orbit potential
on the $pn$ pair around the $^{16}$O  core and discuss the 
parity mixing of the $pn$ pair considering the 
spin-orbit potential from the core as the perturbative external field.
We also discuss the effect of the spin-orbit potential on the four-nucleon correlations, 
i.e., the breaking of the $\alpha$ cluster around $^{16}$O in association with that on the 
$pn$ pair.

This paper is organized as follows. 
In the next section, we describe the present model of the microscopic three-body
$^{16}$O+$p$+$n$ model. 
In \ref{sec:results}, the calculated results for $^{18}$F obtained by the GCM calculations of the $^{16}$O+$p$+$n$ model are shown.
We discuss the behavior of the $pn$ pair focusing on the effect 
of the spin-orbit potential in \ref{sec:discussion}. We also study the four-nucleon 
correlation at the surface of the $^{16}$O core based on the $^{16}$O+$ppnn$ model
in \ref{sec:4-nucleon}.
A summary is given in \ref{sec:summary}. 
In the appendix, 
the mechanism of the breaking of the parity symmetry in the 
pair in the spin-orbit potential is described in appendix \ref{sec:app1}, 
and the mathematical 
relation between the $pn$ cluster wave function and the shell-model wave function
is described in appendix \ref{sec:app2}.

\section{Formulation}\label{sec:formulation}
To investigate the structure of $^{18}$F system focusing 
on the internal structure of the $pn$ pair at the surface around the 
$^{16}$O core,
we adopt a three-body cluster model of $^{16}$O+$p$+$n$ with the form of 
Gaussian wave packets for two valence nucleons, 
\begin{eqnarray}
&&\Phi_{^{16}{\rm O}+pn}(\bvec{R}_1,\bvec{R}_2)  \nonumber\\
&&=  {\cal A} 
\left\{ 
\Phi_{^{16}{\rm O}}\psi_{p\sigma}(\bvec{R}_1) \psi_{n\sigma'}(\bvec{R}_2) \right\},\label{eq:o-pn}\\
&& \psi_{\tau\sigma}(\bvec{R}_j;\bvec{r}_i) 
=  \phi(\bvec{R}_j;\bvec{r}_i) \chi_{\tau\sigma},\\
&&\phi(\bvec{R}_j;\bvec{r}_i)=  \left( \frac{2\nu}{\pi} \right)^{3/4} e^{-\nu(\bvec{r}_i-\bvec{R}_j)^2} \label{eq:single-phi}.
\end{eqnarray}
Here $\Phi_{^{16}{\rm O}}$ is the $^{16}$O  wave function given by the 
H.O. $p$-shell closed configuration with the width $b=1/\sqrt{2\nu}$ fixed 
to be $\nu=0.16$ fm$^{-2}$ to reproduce the root-mean-square radius of $^{16}$O. 
In the practical calculation, $\Phi_{^{16}{\rm O}}$ is approximately written 
by the $4\alpha$ Brink-Bloch wave function with 
the $\alpha$-$\alpha$ distance small enough to express the shell-model limit.
$\chi_{\tau\sigma}$ is the spin-isospin function labeled by $\tau =$ proton or neutron and $\sigma=\uparrow$ or $\downarrow$, and ${\cal A}$ is the antisymmetrizer for all 
nucleons.  The total wave function is the microscopic 18-body wave function.
The spatial parts of single-particle wave functions for valence nucleons are specified by the complex parameters $\bvec{R}_j$ $(j=1,2)$ which stand for the centers of Gaussian wave packets
in the phase space. 
One of the merits of the present model with the wave packet form is that 
the internal wave function and the c.m. motion of the pair are separable, 
\begin{eqnarray}
&&\phi(\bvec{R}_1;\bvec{r}_1)\phi(\bvec{R}_2;\bvec{r}_2) 
= \phi_g(\bvec{R}_g;\bvec{r}_g)  \phi_{\rm in}(\bvec{R};\bvec{r}),\\
&&\phi_g(\bvec{R}_g;\bvec{r}_g)=  \left( \frac{4\nu}{\pi} \right)^{3/4} e^{-2\nu(\bvec{r}_g-\bvec{R}_g)^2}\label{eq:phi_g},\\
&&\phi_{\rm in}(\bvec{R};\bvec{r})=  \left( \frac{\nu}{\pi} \right)^{3/4} e^{-\frac{\nu}{2}(\bvec{r}-\bvec{R})^2}\label{eq:phi_in},\\
&&\bvec{R}_g=\frac{\bvec{R}_1+\bvec{R}_2}{2},\quad
\bvec{R}=\bvec{R}_1-\bvec{R}_2,
\end{eqnarray}
and therefore, the internal structure of the pair can be easily analyzed if we omit the
antisymmetrization effect from the core. 
Moreover, the recoil effect on the core is exactly taken into account when we set the
mean position of the c.m. of $^{16}$O to 
$-\bvec{R}_g/8$. 
Note that the valence nucleon wave function given by the localized Gaussian form
can be expanded by the H.O. shell-model basis around the origin and, 
in the $|\bvec{R}_j|\rightarrow 0$ limit, the single-nucleon wave function after the antisymmetrization 
becomes equivalent to a H.O. $sd$ orbit around the $^{16}$O core
as explained in appendix \ref{sec:app2}.
By transforming the one-center coordinates $\bvec{r}_1$ and $\bvec{r}_2$ to the Jaccobi coordinates 
$\bvec{r}$ and $\bvec{r}_g$, we can switch over from the 
single-particle orbit picture to the $pn$ cluster picture. 

Another merit of the wave packet form is that the mean position and momentum of a 
nucleon are described simply 
by the real part $\bvec{d}_j$ and the imaginary part $\bvec{k}_j/2\nu$ of the Gaussian center parameter 
$\bvec{R}_j=\bvec{d}_j+i \bvec{k}_j/2\nu$ 
($\bvec{d}_j$ and $\bvec{k}_j$ are real vectors), 
\begin{eqnarray}
&&\langle \phi(\bvec{R}_j)| \bvec{r}_i |\phi(\bvec{R}_j)\rangle=  \bvec{d}_j,\\
&&\langle \phi(\bvec{R}_j)| \bvec{p}_i |\phi(\bvec{R}_j)\rangle= \hbar\bvec{k}_j,
\end{eqnarray}
and also those for the c.m. of the $pn$ pair 
\begin{eqnarray}
&& \bvec{R}_g=\bvec{D}_g+i \bvec{K}_g/4\nu, \\
&&\langle \phi_G(\bvec{R}_g)| {\bvec{r}}_{g} |\phi_G(\bvec{R}_g)\rangle=  \bvec{D}_g, \\
&&\langle \phi_G(\bvec{R}_g)| {\bvec{p}}_{g} |\phi_G(\bvec{R}_g)
\rangle = \hbar\bvec{K}_g.
\end{eqnarray}

 In case of $\bvec{R}_1=\bvec{R}_2=\bvec{R}_g$, which corresponds to $\bvec{R}=0$, 
the wave function describes 
the simplest case that two nucleons form the ideal $pn$ cluster  
having the $(0s)^2$ configuration.  
The superposition of the $(0s)^2$  $pn$ cluster wave functions 
with various $\bvec{R}_g$ parameters is equivalent to the 
the generator coordinate method (GCM) of the two-body  $^{16}$O+$(pn)$ cluster model, 
in which $\bvec{R}_g$ is treated as the generator coordinate describing 
the relative motion between the $pn$ cluster and $^{16}{\rm O}$.
Because of the Fermi statistics,
the $(0s)^2$ $pn$ cluster with the isospin $T=0$ 
is the pure spin-triplet ($S=1$) state, while that with $T=1$ is the 
spin-singlet ($S=0$) state. 
In the two-body $^{16}$O+$(pn)$ cluster model, the $T=0$ and $T=1$ states of $^{18}$F are composed 
of the $(TS)=(01)$ and $(10)$ $pn$ clusters, respectively. 

In reality, the ideal $(0s)^2$ $pn$ cluster is broken 
at the nuclear surface because of the effect from the core. Therefore we extend 
the two-body $^{16}$O+$(pn)$ cluster model to the three-body $^{16}$O+$p$+$n$ model 
by taking into account
$\bvec{R}_1\ne\bvec{R}_2$ cases and treat the parameter  $\bvec{R}=\bvec{R}_1-\bvec{R}_2$ as an additional generator coordinate to incorporate more general states of the $pn$ pair having  
internal excitations.
In the present calculation, we perform the GCM calculation of 
the three-body cluster model for $^{18}$F by superposing the 
wave function $\Phi_{^{16}{\rm O}+pn}(\bvec{R}_1,\bvec{R}_2)$. 
Let us consider the body-fixed frame $XYZ$.
We choose the $\bvec{R}_g$ vector on the $Z$-axis as $\bvec{R}_g=(0, 0, D_{g})$ 
and the nucleon-spin ($\sigma$) orientation to the $Y$-axis as  
spin-up ($\uparrow_Y$) or  spin-down ($\downarrow_Y$) in the intrinsic 
wave function before projections. 
In the present GCM calculation for $^{18}$F, 
we restrict the $\bvec{R}$ vector on the $XY$ plane, and treat it 
as the generator coordinate as well as the coordinate $D_{g}$.  
Namely, the parameters chosen in the present GCM calculation are
\begin{eqnarray}
\bvec{R}&=& (\frac{i}{\nu}{k_X},2d_Y,0),\\
\bvec{R}_g&=& (0, 0, D_{g}).
\end{eqnarray}
Here $k_X$, $d_Y$, and $D_{g}$ are real parameters which are treated as 
the generator coordinates. 
Then, the $J^+$ states of $^{18}$F with $T=0$ and $T=1$ are 
given by the linear combination of the 
parity and total-angular-momentum eigen wave functions projected from 
$\Phi_{^{16}{\rm O}+pn}(\bvec{R}_1,\bvec{R}_2)$, 
\begin{eqnarray}
&&\Psi_{^{18}{\rm F}}(T,J^\pi)\nonumber \\
&&=\sum_{k_X,d_Y,D_{g}} c^{T,J^\pi}_{k_X,d_Y,D_{g}}
P^{J\pi}_{MK} \Phi_{^{16}{\rm O}+pn}(\bvec{R}_1,\bvec{R}_2),\\
&&\Phi_{^{16}{\rm O}+pn}(\bvec{R}_1,\bvec{R}_2)\nonumber\\
&&={\cal A} \left\{ 
\Phi_{^{16}{\rm O}}\psi_{p\uparrow_Y}(\bvec{R}_1) 
\psi_{n\downarrow_Y}(\bvec{R}_2) \right\},\\
&&\bvec{R}_1= (i\frac{k_X}{2\nu}, d_Y, D_{g}) \label{eq:r1},\\
&&\bvec{R}_2= (-i\frac{k_X}{2\nu}, -d_Y, D_{g}) \label{eq:r2}.
\end{eqnarray}
Here $P^{J\pi}_{MK}$ is the parity and total-angular-momentum projection operator.
Although each intrinsic wave function contains both $T=0$ and $T=1$ components, 
the isospin symmetry is restored by the $K$ projection, 
and $T=0$ and $T=1$ states
are obtained by projecting onto odd and even $K$ states, respectively.
In the present work, $K=+1$ and $K=-1$ states are mixed for $T=0$ states, 
and $K=0$ is chosen for $T=1$ states by ignoring the isospin breaking in the wave function.
 We omit high $K$ components to save the computational cost 
and avoid the numerical error in the angular-momentum projection. 
The coefficients $c^{T,J^\pi}_{k_X,d_Y,D_{g}}$ are determined by solving the Hill-Wheeler equation 
so as to minimize the energy of $\Psi_{^{18}{\rm F}}(T,J^\pi)$. 

In the three-body cluster GCM calculation, choosing ${\rm Im}[R_X]$ 
as the generator coordinate is equivalent to choosing ${\rm Re}[R_X]$ as the generator 
coordinate. 
We here adopt the imaginary part ${\rm Im}[R_X]=k_X/\nu$ as the generator coordinate
as given in Eqs.~\ref{eq:r1} and \ref{eq:r2}
because it is suitable to consider 
the $\uparrow_Y$ and $\downarrow_Y$ nucleons boosted by the spin-orbit potential at the nuclear surface
to have finite momenta along the $X$ direction. 
Thus we take into account the degrees of $R_X$, $R_Y$ but fix only 
$R_Z=0$ to save the numerical cost. 
The c.m. motion of the $pn$ pair is fully taken 
into account in the present calculation  
with the angular momentum projection and the generator coordinate $D_g$.

To calculate the Gamow-Teller transition strengths from the $^{18}$O to $^{18}$F, 
we also perform 
the GCM calculation of $^{18}$O by using the three-body 
$^{16}$O+$n$+$n$ model in the same way,
\begin{eqnarray}
&&\Psi_{^{18}{\rm O}}(T,J^\pi)\nonumber \\
&&=\sum_{k_X,d_Y,D_{g}} c^{T,J^\pi}_{k_X,d_Y,D_{g}}
P^{J+}_{MK} \Phi_{^{16}{\rm O}+nn}(\bvec{R}_1,\bvec{R}_2),\\
&&\Phi_{^{16}{\rm O}+nn}(\bvec{R}_1,\bvec{R}_2)\nonumber\\
&&={\cal A} \left\{ 
\Phi_{^{16}{\rm O}}\psi_{n\uparrow_Y}(\bvec{R}_1) 
\psi_{n\downarrow_Y}(\bvec{R}_2) \right\},\\
&&\bvec{R}_1= (i\frac{k_X}{2\nu}, d_Y, D_{g}), \\
&&\bvec{R}_2= (-i\frac{k_X}{2\nu}, -d_Y, D_{g}).
\end{eqnarray}

In a similar way to the $^{16}$O+$p$+$n$ wave function, we also consider the $^{16}$O+$4N$ model 
to study the $\alpha$ cluster breaking at the surface of the $^{16}$O core,
\begin{eqnarray}
&&\Phi_{^{16}{\rm O}+4N}(\bvec{R}_1,\bvec{R}_2,\bvec{R}_3,\bvec{R}_4)  \nonumber\\
&&=  {\cal A} 
\left\{ 
\Phi_{^{16}{\rm O}}\psi_{p\uparrow_Y}(\bvec{R}_1) \psi_{p\downarrow_Y}(\bvec{R}_2) 
\psi_{n\uparrow_Y}(\bvec{R}_3) \psi_{n\downarrow_Y}(\bvec{R}_4)  
 \right\}. \nonumber\\
\label{eq:o16-4N}
\end{eqnarray}

\section{Results}\label{sec:results}

\subsection{Effective nuclear force}
The present model is based on the microscopic $A$-body wave function
with the assumption of the inert core.
In the model, the Hamiltonian consists of the one-body kinetic term and the 
effective two-body nuclear forces and Coulomb force,
\begin{equation}
H=\sum_{i} t_i - T_G + \sum_{i<j} v^{\rm eff}_{ij} + \sum_{i<j} v^{\rm Coul}_{i,j}.
\end{equation}
Here $T_G$ is the kinetic energy of the total c.m. motion. 
We adopt the Volkov No.2 force \cite{VOLKOV} for the effective central forces and supplement the spin-orbit force
with the same form as the spin-orbit term of the G3RS force \cite{LS}.
Then, the central and spin-orbit forces are expressed by two-range Gaussian forms,
\begin{eqnarray}
&& v^{\rm eff}_{ij} \nonumber\\ 
&& =  v_{\rm c}(r_{ij}) \left( (1-m)+b P^\sigma_{ij}-h P^\tau_{ij} -m P^\sigma_{ij} P^\tau_{ij} \right)
\nonumber\\
&& +  v_{\rm ls}(r_{ij}) P(^3{\rm O}) \left( \bvec{l}_{ij}\cdot \bvec{S}_{ij} \right), \\
&&r_{ij}=|\bvec{r}_i-\bvec{r}_j|,\\
&& v_{\rm c}(r) =\sum_{k=1,2}
v^c_{k} e^{-\frac{r^2}{a^2_{k}}},\\
&& v_{\rm ls}(r_{ij})=\sum_{k=1,2}
v^{\rm ls}_{k} e^{-\frac{r^2}{a'^2_{k}}},
\end{eqnarray}
where $P^\sigma_{ij}$ and $P^\tau_{ij}$ are the spin and isospin exchange operators, $P(^3{\rm O})$ is the projection 
operator onto the triplet odd ($^3{\rm O}$) state,  
$\bvec{l}_{ij}$ is the angular momentum for 
the relative coordinate $\bvec{r}_{ij}\equiv \bvec{r}_i-\bvec{r}_j$, and $\bvec{S}_{ij}$ 
is the sum of the nucleon spins
$\bvec{S}_{ij}=\bvec{s}_i+\bvec{s}_j$. 
The constant values $v^c_{1,2}$ and $a_{1,2}$ are the strength 
and range parameters of the central force given in the Volkov No.2 force, 
and $v^{\rm ls}_{1,2}$ and $a'_{1,2}$ are those of the spin-orbit force
in G3RS.

We choose the Majorana parameter $m=0.62$
which is the same parameter used in Ref.~\cite{Matsuse75} to 
reproduce the $^{20}$Ne energy spectra measured from the threshold energy in 
the $^{16}$O+$\alpha$ cluster model calculation.
The  $b$ and $h$ for Bartlett and Heisenberg terms are the adjustable parameters
that change the relative strength 
of the nuclear forces in the $^3$E and the $^1$E channels. 
In the present calculation, we mainly use $b=h=0.125$ which reproduce the 
deuteron binding energy in the $^3$E channel
as well as the unbound feature of two nucleons in the $^1$E channel. 
The ratio $f$ of the $S$-wave $NN$ force in the $^3$E channel to that in the $^1$E channel
is $f=(1+b+h)/(1-b-h)=1.67$ for $b=h=0.125$.
Although those parameters describe reasonably the properties of two-nucleon systems in free space, 
they are found to overestimate excitation energies of the $T=1$ states relative to the 
$T=0$ states in $^{18}$F. Therefore, we also demonstrate the results calculated with 
the parameter set, $b=h=0.06$ with the weaker ratio $f=1.23$.
We use the labels "bh125" and "bh06" for the interaction parameters $b=h=0.125$ and
$b=h=0.06$, respectively. 

\begin{figure}[tb]
\begin{center}
\includegraphics[width=6.5cm]{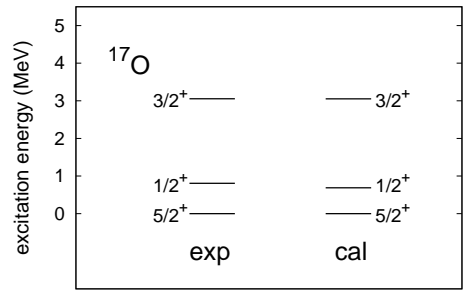} 	
\end{center}
\vspace{0.5cm}
  \caption{Energy levels of $^{17}$O. Theoretical energies are calculated 
by the GCM calculation of two-body $^{16}{\rm O}+n$ model using the bh125 interaction with the spin-orbit force
with the strength $u_{\rm ls}=820$ MeV.
\label{fig:o17-spe}}
\end{figure}

For the strength of the two-body spin-orbit force, 
$v^{\rm ls}_{1}=-v^{\rm ls}_{2}\equiv u_{\rm ls}=820$ MeV 
is chosen to reproduce the low-lying energy spectra of $^{17}$O in the 
$^{16}{\rm O}+n$ model calculation (see Fig.~\ref{fig:o17-spe}).
To discuss the effect of the spin-orbit force on the $pn$ pair in $^{18}$F, we 
perform calculations with and without the spin-orbit force. 

For the two-body Coulomb force 
\begin{equation}
v^{\rm Coul}_{i,j}= \sum_{i<j} \frac{1+P^\tau_i}{2}\frac{1+P^\tau_j}{2} \frac{e^2}{r_{ij}}, 
\end{equation}
the function $1/r$ is approximated by a sum of seven Gaussians. 

\subsection{Energy levels of $^{18}$F}
We perform the GCM calculation of the three-body $^{16}$O+$p$+$n$ model. 
The generator coordinates, $D_{g}$, $k_X$, and $d_Y$  are discretized as
$D_{g}=1,2,3,4,5$ fm, $k_X/2\nu=0, 0.5, \cdots, a^{\rm max}_X$ fm with 
$a^{\rm max}_X=\min (1+D_{g}, 2.5 {\rm\ fm})$, and 
$d_Y=0,1,2$ fm. We check the convergence for the maximum values of these parameters, and it is found that, 
when basis wave functions with the maximum values are excluded, the change of the excitation 
energy is less than 0.1 MeV.

\begin{figure}[tb]
\begin{center}
\includegraphics[width=6.5cm]{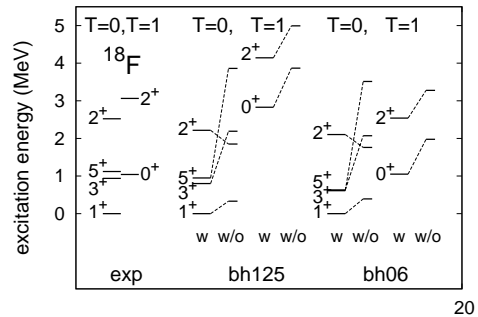} 	
\end{center}
\vspace{0.5cm}
  \caption{Energy levels of $T=0$ $J^\pi=1^+$, $2^+$, $3^+$, and $5^+$ states, and $T=1$ $J^\pi=0~+$ and $2^+$ states in $^{18}$F obtained with the GCM calculation of the three-body $^{16}{\rm O}$+$p$+$n$ model in comparison with the  
experimental levels for the corresponding states. 
The results obtained with the bh125 and bh06 interactions are shown. 
The energies without the spin-orbit force measured from the $T=0$ $J^\pi=1^+$ 
energy obtained with the spin-orbit force are also shown. 
\label{fig:18F-spe}}
\end{figure}

The calculated energy levels of $^{18}$F are shown in Fig.~\ref{fig:18F-spe}
compared with the experimental data.
In the result calculated with the spin-orbit force, 
we obtain $J^\pi=1^+$, $3^+$, $5^+$ states, as well as the $2^+$ state in the $T=0$ channel
and $J^\pi=0^+$ and $2^+$ states in the $T=1$ channel. The calculated
energy spectra in each $T=0$ and $T=1$ channel are in good agreement with the experimental data.
However, in the result with the bh125 interaction, the relative energies of $T=1$ states to the $T=0$ states are overestimated compared with the experimental data. 
In the calculation with the bh06 interaction 
the experimental energy spectra of $T=0$ and $T=1$ states are reproduced reasonably. 
As mentioned before, the bh125 interaction is a reasonable effective nuclear force for 
$S$-wave two nucleons in free space. In the present three-body model
for $^{18}$F, we need to empirically modify the $b$ and $h$ parameters 
to reproduce the relative energy between 
$T=0$ and $T=1$ states. 
We do not know the fundamental reason for the modification, but 
it may originate in the limitation of the effective two-body interaction in the microscopic calculation 
with the inert $^{16}$O core assumption. 
It is found that structures of the ground and excited states do not depend so much 
on the choice of the bh125 and bh06 interactions except for the $T=0$ and $T=1$ relative energy. Therefore, in this paper, we mainly discuss the results calculated 
with the original bh125 interaction. 

We also show the energy spectra calculated by switching off the spin-orbit force.
The energies are measured from the $J^\pi=1^+$ ground state energy obtained with the spin-orbit force. 
In the energy spectra without the spin-orbit force, 
the intrinsic spin $S$ decouples from the orbital-angular momentum $L$.
As a result,
the $T=0$ levels can be understood by the simple $L=0$, 2, and 4 spectra for 
$J=L\pm 1$ states in the $LS$ coupling scheme.
In the present calculation, slight coupling of $L$ and $S$ 
remains because of the truncation of the 
model space. 

Comparing the spectra with and without the spin-orbit force, it is found that 
the $T=0$ spectra are drastically changed by the spin-orbit force.
The energy gain due to the spin-orbit force is very small in the $J^\pi=1^+$ state, while 
that in the $3^+$ state is as large as about 1 MeV, and it is largest 
in the $5^+$ state as about 3 MeV. 
Naively, the intrinsic spin $S=1$ and 
the orbital angular momentum $L$ of the $(TS)=(01)$ $pn$ pair 
are parallel in the $J=L+1$ states and 
two nucleons in the $pn$ pair in finite $L$ states feel the 
attractive spin-orbit potential (see upper panels of Fig.~\ref{fig:pn-pair})
but the pair in the $L=0$ state feels no spin-orbit potential.
It means no energy gain in the $L=0$ state and the larger gain in high $L$ states.
Indeed, the systematics of the energy gain in the present result follows this rule in principle. 

In the $T=1$ energy spectra, 
the energy gain due to the spin-orbit force
is about 1 MeV in the $J^\pi=0^+$ state, and it is slightly smaller in the $2^+$
state. The $0^+$-$2^+$ level spacing in the $T=1$ spectra does not change so much 
compared with the drastic change in the $T=0$ spectra by the spin-orbit force.
The spin-orbit potential energy gain of the $T=1$ $pn$ pair is caused by boosting the spin-up and -down nucleons in the opposite direction
(see upper panels of Fig.~\ref{fig:pn-pair}).
Because of the larger energy gain with the spin-orbit force in the $T=1$ $J^\pi=0^+$ state than the
$T=0$ $J^\pi=1^+$ state, 
the excitation energy of the $T=1$ $J^\pi=0^+$ state becomes lower
than the case without the spin-orbit force. 
This result is consistent with the studies of the $T=0$ pairing 
based on shell-model calculations and mean-field calculations in preceding works 
\cite{Poves:1998br,Baroni:2009eh,Bertsch:2009xz,Gezerlis:2011rh,Sagawa:2012ta,Bertsch:2009zd}
that suggested the unlikely $T=0$ $J^\pi=1^+$ pair
compared with the favored $T=1$ $J^\pi=0^+$ pair
in the spin-orbit potential. 

It should be noted that the $(TS)=(01)$ pair in high $L$ states around 
the core is favored by 
the spin-orbit field. As a result, the high spin states with $T=0$ come down 
to the low-energy region.  
Detailed discussions of the energy gain mechanism 
of $T=0$ and $T=1$ $pn$ pairs in the spin-orbit potential are given in the next section.

\subsection{Spin components in $^{18}$F}

If there is no spin-orbit potential, the $T=0$ and $T=1$ $pn$ pairs around the $^{16}$O core
have the pure $S=1$ and $S=0$ components, respectively. 
However, in the spin-orbit potential, the $S=0$ and $S=1$ components mix in 
the $T=0$ and $T=1$ pairs, respectively.
The mixing of different intrinsic spin components corresponds to 
the odd-parity mixing in the pair caused by the spin-orbit field from the $^{16}$O core.
According to Fermi statistics of nucleons,
the mixed odd-parity component ${\cal P}_{\rm odd}$ in the pair can be measured by 
the expectation values of the squared intrinsic spin $\langle \bvec{S}\rangle ^2$ 
as 
${\cal P}_{\rm odd}=1-\langle \bvec{S}^2 \rangle/2$ for the $T=0$ states and 
${\cal P}_{\rm odd}=\langle \bvec{S}^2 \rangle/2$ for the $T=1$ states.

The calculated values of $\langle \bvec{S}^2\rangle$ and ${\cal P}_{\rm odd}$
for the $^{18}$F states obtained by the GCM calculation are shown 
in Table \ref{tab:18F-spin}. 
In the results obtained with the spin-orbit force, 
the significant odd-parity ($S=1$) mixing is found in 
the $T=1$ states. On the other hand, the odd-parity mixing ($S=0$) is smaller in the $T=0$ states,
in particular, in high spin states.
In other words, the parity mixing occurs in the $T=1$ pair, 
while the parity mixing is suppressed in the $T=0$ pair. 
It indicates that the spin-orbit potential breaks the $(TS)=(10)$ pair more than the
$(TS)=(01)$ pair. It is simply understood by the role of the spin-orbit potential boosting 
two nucleons in the pair because 
spin anti-parallel two nucleons in the $S=0$ pair 
feel the momentum dependent force in the opposite direction, 
while spin parallel two nucleons in the $S=1$ pair feel
the momentum dependent force in the same direction. 
We give detailed discussions later. 

\begin{table}[ht]
\caption{
\label{tab:18F-spin} 
The expectation values $\langle \bvec{S}^2 \rangle$ of the squared intrinsic spin 
of the $T=0$ and $T=1$ states of
$^{18}{\rm F}$ obtained by the GCM calculations of the three-body $^{16}$O+$p$+$n$ model. 
The results calculated with the bh125 and the bh06 interactions 
with the spin-orbit force, and the results of the bh125 force without 
the spin-orbit force are listed.
The odd-parity component ${\cal P}_{\rm odd}$ 
in the $pn$ pair for each state is also shown. 
}
\begin{center}
\begin{tabular}{|c|cc|cc|cc|}
&\multicolumn{2}{|c|}{ bh125 }&
\multicolumn{2}{|c|}{ bh06 }&
\multicolumn{2}{|c|}{ bh125 w/o ls }\\
$J^\pi$, $T$&  $\langle \bvec{S}^2 \rangle$ & ${\cal P}_{\rm odd} $
& $\langle \bvec{S}^2 \rangle$ & ${\cal P}_{\rm odd} $
&  $\langle \bvec{S}^2 \rangle$ & ${\cal P}_{\rm odd}$ \\
$0^+$, $T=1$ & 0.23 & 0.11 &	0.19 &	0.09 &	0.00 &	0.00\\ 
$2^+$, $T=1$	&0.27 &	0.14 &	0.21 &	0.11 &	0.00 &	0.00 \\
$1^+$, $T=0$	&1.91 &	0.05 &	1.89 &	0.06 &	2.00 &	0.00 \\
$3^+$, $T=0$	&1.95 &	0.02 &	1.94 &	0.03 &	2.00 &	0.00 \\
$5^+$, $T=0$	&2.00 &	0.00 &	2.00 &	0.00 &	2.00 &	0.00 \\
$2^+$, $T=0$	&1.95 &	0.03 &	1.93 &	0.04 &	2.00 &	0.00 \\
\end{tabular}
\end{center}
\end{table}

\subsection{$M1$ and GT Transitions of $^{18}$F}
To check the reliability of spin configurations in the present calculation, 
we show the calculated values of the magnetic moments, the $M1$ transition strength, and 
the GT transition strength of $^{18}$F and compare them with the experimental data in Table \ref{tab:18F-transition}.
For $B$(GT), we perform the GCM calculation of the $^{16}$O+$n$+$n$ model to obtain 
the $^{18}$O ground state in the same way as that in the $^{16}$O+$p$+$n$ model
for $^{18}$F.
The calculated results
are in reasonable agreement with the experimental data.

As discussed before, the dominant components of 
the $T=0$ $J^\pi=1^+$, $3^+$, and $5^+$ states
are the $S=1$ states coupling with the orbital angular 
momentum $L=0$, 2, and 4, and the mixing of $S=0$ component is minor.
Therefore, the magnetic moments of these states
are not sensitive to the interaction
because the intrinsic spin configurations 
do not depend so much on the interaction. 

The results of $B(M1)$ for $^{18}{\rm F}(0^+)\rightarrow^{18}{\rm F}(1^+)$ and 
$B({\rm GT})$ for $^{18}{\rm O}(0^+)\rightarrow ^{18}{\rm F}(1^+)$
obtained
using the bh125 interactions with the spin-orbit force 
are similar to those using the bh06 interaction with the spin-orbit force. 
However, 
they are somewhat different from the result obtained without the spin-orbit force
because the spin structure changes significantly in the $T=1$ $J^\pi=0^+$ state 
reflecting the $S=1$ mixing, i.e., the parity mixing in the $T=1$ pair.
In the result without the spin-orbit force, $B({\rm GT})$ is remarkably large showing that the 
transition is the super allowed transition given by the spin-isospin flip from 
the $S=1$ $pn$ pair in $^{18}{\rm F}(1^+)$ to the 
$S=0$ $nn$ pair in $^{18}{\rm O}(0^+)$. With the spin-orbit force,
$B({\rm GT})$ becomes small because the spin structure of 
$nn$ pair in $^{18}{\rm O}(0^+)$ is changed by the spin-orbit force.

\begin{table}[ht]
\caption{
\label{tab:18F-transition} 
The magnetic moments of $^{18}{\rm F}(1^+)$, $^{18}{\rm F}(3^+)$, and 
$^{18}{\rm F}(5^+)$, 
the $B(M1)$ for the transition $^{18}{\rm F}(0^+)\rightarrow^{18}{\rm F}(1^+)$, 
and the $B({\rm GT})$ for the transition 
$^{18}{\rm O}(0^+)\rightarrow ^{18}{\rm F}(1^+)$. The results calculated with the bh125 and bh06 interactions with the spin-orbit force, and the bh125 interaction without 
the spin-orbit force are shown. The experimental data are taken from Refs.~\cite{stone05,tilley95}.
}
\begin{center}
\begin{tabular}{|c|cccc|}
& exp. & bh125 & bh06 & w/o ls \\
$\mu(1^+)$ ($\mu_N$) & $-$ & 0.82 &	0.82 &	0.82 \\ 
$\mu(3^+)$ ($\mu_N$) & 1.77(12)	& 1.86 &	1.85 &	1.84 \\
$\mu(5^+)$ ($\mu_N$) & 2.86(3)&	2.88 &	2.88 &	2.88 \\
$B(M1;0^+)$ ($\mu_N^2$) &	19.5(3.8)&	17.0 &	17.4 &	14.1 \\ 
$B({\rm GT})$ ($g_A^2/4\pi$)
& 3.18	& 5.0 &	5.2 &	10.7 \\
\end{tabular}
\end{center}
\end{table}

\section{Discussion}\label{sec:discussion}
We discuss here the effect of the spin-orbit potential on the $pn$ pair at the nuclear surface, 
in particular, the effect on the internal structure of the pair
and that on the c.m. motion of the pair. We show that 
the symmetry breaking, i.e., the odd-parity mixing in the pair occurs due to 
the spin-orbit field from the core nucleus. 

\subsection{Basic idea of effect of spin-orbit field on a $pn$ pair at nuclear surface}

As mentioned before, 
the $^{18}$F energy spectra obtained without the spin-orbit force
can be understood by the dinucleon picture that the 
$(TS)=(10)$ pair or the $(TS)=(01)$  pair is moving 
in the $L$ wave around the $^{16}$O core. 
In the case with the spin-orbit force, 
the odd-parity mixing occurs and the $T=0$ and $T=1$ pairs are no longer the pure
$(TS)=(10)$ and $(TS)=(01)$ states.
In the low spin $J^\pi=0^+$ and $1^+$ states of the pair dominantly in the lowest $L=0$ mode, 
the $T=1$ pair gains the spin-orbit potential energy 
while involving the odd-parity ($S=1$) mixing in the dominant 
even-parity ($S=0$) component, but the $T=0$ pair does not gain the
spin-orbit potential energy so much and it has
only minor odd-parity mixing. 
On the other hand, the $T=0$ pair in high $L$ states is 
favored largely by the spin-orbit field without the odd-parity mixing in the pair. 

To understand features of the $pn$ pair around the $^{16}$O core
we consider a single wave function $\Phi_{^{16}{\rm O}+pn}(\bvec{R}_1,\bvec{R}_1)$
which expresses a $pn$ pair localized around a certain position $\bvec{R}_g$ 
at the surface of the core in the body-fixed $XYZ$ frame defined 
in Sec.\ref{sec:formulation}.
Starting from the ideal dinucleon of the $(0s)^2$ $pn$ pair in the case without the spin-orbit field
and considering the spin-orbit force as the perturbative external field from the core, 
we discuss how the pair behavior is affected by the spin-orbit field. 
The ideal $(0s)^2$ $pn$ pair localized around the position $(0,0,D_g)$
is written by $\Phi_{^{16}{\rm O}+pn}(\bvec{R}_1,\bvec{R}_2)$ with
$\bvec{R}_1=\bvec{R}_2=(0,0,D_{g})$. 
We consider the spin parallel pair consisting of a $\uparrow_Y$ proton and a $\uparrow_Y$ 
neutron and the spin anti-parallel pair of 
a $\uparrow_Y$ proton and a $\downarrow_Y$ neutron in the intrinsic frame $XYZ$. 
The former is the $(TS)=(01)$ state, and the latter contains 
the $T=0$ and $T=1$ components which are decomposed by the $J_Z=K$ projection.
Firstly we discuss the pair behavior in the intrinsic frame, and later 
we discuss the decomposition of the $\uparrow_Y\downarrow_Y$ pair into 
$T=1$ and $T=0$ components.

When the spin-orbit force is switched on, nucleons in the $\uparrow_Y\uparrow_Y$
pair are boosted by the one-body spin-orbit field from the $^{16}$O core
to have finite momentum in the same direction along the $X$-axis, while 
$\uparrow_Y$ and $\downarrow_Y$ nucleons in the spin anti-parallel pair 
are boosted in the opposite direction to each other as shown in Fig.\ref{fig:pn-pair}. 
The boosting mechanism by the spin-orbit potential can be easily understood 
by the following classical picture of the Gaussian wave packet for each nucleon
moving on the $XY$ plane passing through the $(0,0,D_{g})$.
The nucleon at the position $(0,0,D_{g})$ having the finite momentum
$\bvec{k}_i=(k_{jX},k_{jY},0)$ has the angular momentum 
$\bvec{l}_i=D_{g}(-k_{jY},k_{jX},0)$.
Assuming the averaged one-body spin-orbit potential
around the position $(0,0,D_{g})$ as $-\bar{U}_{ls}\bvec{l}_i\cdot \bvec{s}_i$ with 
a constant positive value $\bar{U}_{ls}$, the spin-orbit potential energy gain 
for the $\uparrow_Y$ nucleon is given as $-\frac{\bar{U}_{ls}D_{g}}{2}k_{jX}$ and 
that for the $\downarrow_Y$ nucleon is $\frac{\bar{U}_{ls}D_{g}}{2}k_{jX}$. 
It means that the spin-orbit potential energy gain is proportional to 
$k_{jX}$ (nucleon momentum along the $X$ axis) and the sign of $k_{jX}$ should be positive 
for the $\uparrow_Y$ nucleon and it should be negative for the $\downarrow_Y$ nucleon. 

As for the $\uparrow_Y\uparrow_Y$ pair for the $T=0$ pair, the spin-orbit potential 
boosts two nucleons along the $X$ axis in the same direction to excite 
the rotational mode of the c.m. motion of the pair 
keeping the internal pair wave function unchanged.
In the $\Phi_{^{16}{\rm O}+pn}(\bvec{R}_1,\bvec{R}_2)$ model wave function, 
the state with two $\uparrow_Y$ nucleons boosted in the same direction is described 
by the parameter set $\bvec{R}_1=\bvec{R}_2=(i\kappa_X/2\nu,0,D_{g})$. In such a state, 
$\bvec{R}=\bvec{R}_1-\bvec{R}_2$ equals zero and
the internal pair wave function $\phi_{\rm in}(\bvec{R};\bvec{r})$ in Eq.~\ref{eq:phi_in} 
is consistent with the $(0s)^2$ $pn$ pair, while the c.m. wave function 
of the pair $\phi_g(\bvec{R}_g;\bvec{r}_g)$ in Eq.~\ref{eq:phi_g} with 
$\bvec{R}_g=(i\kappa_X/2\nu,0,D_{g})$ indicates the $pn$ pair localized around the position 
$\bvec{D}_g=(0,0,D_{g})$ with the finite momentum $\bvec{K}_g=(2\kappa_X,0,0)$.

On the other hand, two nucleons in the $\uparrow_Y\downarrow_Y$ pair 
are boosted by the spin-orbit potential in the opposite direction to each other. 
The momenta of two nucleons in the opposite direction inevitably causes 
the $P$-wave mixing to the dominant $S$-wave component in the internal wave function of the pair, 
and it is regarded as the parity mixing, a kind of symmetry breaking of the pair as follows.
As shown later, 
in the $\Phi_{^{16}{\rm O}+pn}(\bvec{R}_1,\bvec{R}_2)$ model space with the constraint
$R_{jZ}=D_g$, the energy minimum solution of the $\uparrow_Y\downarrow_Y$ pair 
in the intrinsic frame is given by the parameter set  
$\bvec{R}_1=(ik_X/2\nu,0,D_{g})$ and $\bvec{R}_2=(-ik_X/2\nu,0,D_{g})$ 
corresponding to the $\uparrow_Y$ proton and the 
$\downarrow_Y$ neutron boosted in the opposite direction.
The $\uparrow_Y\downarrow_Y$ pair with the finite $k_X$ 
is no longer the $(0s)^2$ pair, but it 
contains the odd-parity mixing in the dominant even-parity component
in the internal pair wave function. 
Indeed the corresponding internal wave function $\phi_{\rm in}(\bvec{R}; {\bf r})$ of the pair for $\bvec{R}=\bvec{R}_1-\bvec{R}_2=(ik_X/\nu,0,0)$ is given as
\begin{eqnarray}
\phi_{\rm in}(\bvec{R};\bvec{r})=  \left( \frac{\nu}{\pi} \right)^{3/4}
e^{-\frac{\nu}{2}(x-ik_X/\nu)^2-\frac{\nu}{2}y^2-\frac{\nu}{2}z^2},
\end{eqnarray}
and its Taylor expansion with respect to $k_X$ is 
\begin{eqnarray}
&&\phi_{\rm in}(\bvec{R};\bvec{r}) \propto \left( \frac{\nu}{\pi} \right)^{3/4}
\left(1+ik_X x + \mathcal{O}(k^2_X)\right) e^{-\frac{\nu}{2}\bvec{r}^2}\nonumber\\
&&=(0,0,0)_{\rm ho}+i\sqrt{\frac{1}{2\nu}}k_X(1,0,0)_{\rm ho}+ \mathcal{O}(k^2_X).
\end{eqnarray}
Here  $(n_X,n_Y,n_Z)_{\rm ho}$ is the H.O. solution for the width $b^2=1/\nu$. 
As clearly seen, the $P$-wave $(1,0,0)_{\rm ho}$ component mixes in the $S$-wave $(0,0,0)_{\rm ho}$ state.
Namely, the parity mixing, i.e., the symmetry breaking in the $\uparrow_Y\downarrow_Y$ pair occurs 
because of the external spin-orbit field. Here, the parameter $k_X$ is regarded as 
the order parameter for the symmetry breaking.
The energy gain is the second order perturbation caused by the 
transition from the $S$-wave state to the $P$-wave state. 
The parity mixing mechanism in the dinucleon pair is also explained in appendix \ref{sec:app1}.

It should be commented that a similar phenomenon of the parity mixing of pairs 
has been discussed recently in the condensed matter physics \cite{Fujimoto09}.
The parity mixing of the Cooper pairs was suggested to occur because of the spin-orbit interaction 
in noncentrosymmetric superconductors having the breaking of 
inversion symmetry in the crystal structure,  
The mechanism of the symmetry breaking is analogous to the above-mentioned
parity mixing in the two-nucleon pair at the nuclear surface due to the spin-orbit field
from the core.

Needless to say, the parity and the rotational symmetries in the total system are restored 
in energy eigen states in the laboratory frame.
In the present model, it is realized by the parity and total angular momentum projection.
We consider the correspondence of the $pn$ pair in the intrinsic frame to the $^{18}$F states
in the laboratory frame. 

For the $\uparrow_Y\uparrow_Y$ pair,
the spin-orbit field causes the rotational excitation of the
c.m. motion of the pair in the body-fixed frame as explained before. 
In the laboratory frame, it corresponds to the $(TS)=(01)$ pair 
in the high orbital angular momentum $L$ state around the core. 
In other words, the $(TS)=(01)$ pair in high $L$ states is favored by 
the spin-orbit field. This is a naive explanation for 
the larger energy gain of the spin-orbit in the higher spin states in the $T=0$ spectra
of $^{18}$F found in the GCM calculation in the previous section. 

On the other hand, for the $\uparrow_Y\downarrow_Y$ pair, the spin-orbit potential
chances the internal structure of the pair involving the odd-parity mixing. However, 
the c.m. motion of the pair in the body-fixed frame is not affected by the spin-orbit force. 
In the laboratory frame, it leads to 
less sensitivity of the 
the $0^+$-$2^+$ level spacing in the $T=1$ spectra of $^{18}$F on the spin-orbit force
as shown in the GCM calculation.

The $\uparrow_Y\downarrow_Y$ pair localized around $(0,0,D_g)$ 
contains the $T=1$ and $T=0$ components
which are decomposed by the $K$ projection. 
In the intrinsic frame, 
the spin-orbit potential energy gain 
of the $\uparrow_Y\downarrow_Y$ pair is proportional to 
the odd-parity mixing in the pair and it is mainly given by the non-diagonal matrix elements of 
the spin-orbit potential between the $S$-wave and $P$-wave states in the pair.
However, since the odd-parity mixing inevitably causes the internal energy loss of the pair, 
the mixing ratio of the odd-parity component to the even-parity component in the 
$\uparrow_Y\downarrow_Y$ pair is determined by 
the competition between the spin-orbit potential energy gain and the internal energy loss. 
  
In case of the high spin $J^\pi=3^+$ and $5^+$ states, 
the $\uparrow_Y\uparrow_Y$ pair moving in $L=2$ and $L=4$ wave is naively 
expected to be favored because it gains the spin-orbit potential energy
without the internal energy loss. 
For the $J^\pi=1^+$ state, we can consider
the $T=0$ component projected from the $\uparrow_Y\downarrow_Y$ pair
instead of the $\uparrow_Y\uparrow_Y$ pair
because there is no energy gain for the $\uparrow_Y\uparrow_Y$ pair in the $L=0$ wave.
However, in the $T=0$ component of the $\uparrow_Y\downarrow_Y$ pair, 
the odd-parity mixing in the pair is unfavored because it suffers from the larger internal energy 
loss compared with the $T=1$ component of the $\uparrow_Y\downarrow_Y$ pair
because of the following reason.
The $T=1$ and $T=0$ components of the $\uparrow_Y\downarrow_Y$ pair are 
obtained by the $K=0$ and $K=1$ projections, respectively.
For a given finite value of $k_X$, the odd-parity mixing becomes half of 
the intrinsic state in the $K=0$ projected state for the $T=1$ component 
but it is not the case in the $K=1$ state for the $T=0$ component 
as explained in appendix \ref{sec:app1}. 
As a result, the internal energy loss of the pair is less 
in the $T=1$ pair than the $T=0$ pair, and 
therefore the $T=1$ pair efficiently gains the spin-orbit potential 
involving the parity mixing. 
In other words, the spin-orbit potential energy gain with the parity mixing is not so 
efficient for the $T=0$ pair in the $J^\pi=1^+$ state as the $T=1$ pair in the $J^\pi=0^+$ state. 
More quantitative discussion is given in the later analysis.  

This mechanism of the unfavorable $T=0$ $J^\pi=1^\pi$ pair 
in the spin-orbit potential is consistent with that argued by Bertsch \cite{Bertsch:2009zd}.
It is also consistent with the discussions in 
Refs.~\cite{Poves:1998br,Baroni:2009eh,Bertsch:2009xz,Gezerlis:2011rh,Sagawa:2012ta},
where the relation between the $jj$ coupling and the $LS$ coupling schemes was 
discussed.
One of the new standpoints in the present dinucleon picture is that 
we focus on the internal pair wave function and discuss its change 
involving the odd-parity mixing because of the spin-orbit force. 
That is to say, starting from the $S=0$ pair with the pure even-parity 
component in the case without the spin-orbit field,  
we consider the odd-parity ($S=1$) mixing in the pair 
caused by the perturbative spin-orbit field. This is an alternative 
interpretation of the mean field picture in the $jj$ coupling scheme 
where even-parity ($S=0$) and odd-parity ($S=1$) components are already mixed 
in the $j^2$ state in no correlation limit and the enhancement of the 
even-parity component by the pair correlation is taken into account. 

Generally, in $J^\pi$ states after the parity and angular momentum projection, 
internal degrees of freedom in the pair and the c.m. motion of the pair 
are not separable. However, according to Fermi statistics of nucleons, 
we can measure the mixing of the odd-parity components 
with the spin-singlet and spin-singlet components in the $^{18}$F states 
obtained by the GCM calculation. As discussed in the previous section, 
it is found in the GCM calculation that 
the odd-parity mixing is significant in the $T=1$ states while it is minor in the 
$T=0$ states consistently with the above naive explanation. 
We should comment that 
the odd-parity component in the pair couples with the 
odd-parity wave of the pair c.m. motion around the core in 
the $J^+$ states in the laboratory frame.


\subsection{Analysis based on a single $^{16}$O+$pn$ cluster wave function}

As mentioned, the spin-orbit force changes the internal structure of the 
$\uparrow_Y\downarrow_Y$ pair and the c.m. motion of 
the $\uparrow_Y\uparrow_Y$ pair in the body-fixed frame. 
We here demonstrate how the parity mixing of the $T=1$ pair occurs in the 
microscopic $^{18}$F system based on two-body effective nuclear forces.
We analyze the energy expectation value of a single three-body 
$^{16}$O+$p$+$n$ wave function
$\Phi_{^{16}{\rm O}+pn}(\bvec{R}_1,\bvec{R}_2)$
and quantitatively discuss 
the effect of the spin-orbit force on the $T=1$ and the $T=0$ pairs.

In the wave function
$\Phi_{^{16}{\rm O}+pn}(\bvec{R}_1,\bvec{R}_2)$, 
single-nucleon wave functions for two nucleons in the $pn$ pair are specified by 
the complex parameters $\bvec{R}_1$ and $\bvec{R}_2$ which express the centers of 
single-nucleon Gaussian wave packets in the phase space. 
To see the behavior of the $pn$ pair localized around 
a certain position $\bvec{R_g}=(\bvec{R}_1+\bvec{R}_2)/2=(0,0,D_{g})$
with the distance $D_g$ from the core, 
we vary the parameters $\bvec{R}_1$ and $\bvec{R}_2$ on the $XY$ plane 
passing through the $(0,0,D_{g})$ and 
obtain the energy minimum solution under the constraints $R_{1X}=-R_{2X}$, $R_{1Y}=-R_{2Y}$, 
and $R_{1Z}=R_{1Z}=D_{g}$.

We first perform the energy variation with respect to the intrinsic state
$\Phi_{^{16}{\rm O}+pn}(\bvec{R}_1,\bvec{R}_2)$
without the parity and the total-angular-momentum projections,
\begin{equation}
\delta \frac{\langle \Phi_{^{16}{\rm O}+pn}(\bvec{R}_1,\bvec{R}_2)|H|\Phi_{^{16}{\rm O}+pn}(\bvec{R}_1,\bvec{R}_2) \rangle}
{\langle \Phi_{^{16}{\rm O}+pn}(\bvec{R}_1,\bvec{R}_2)|\Phi_{^{16}{\rm O}+pn}(\bvec{R}_1,\bvec{R}_2) \rangle}=0.
\end{equation}
The $\uparrow_Y\downarrow_Y$ and $\uparrow_Y\uparrow_Y$ pairs are considered. 
The former contains the $T=0$ and $T=1$ components, while 
the latter is regarded as the $T=0$ pair.
In the total angular momentum projection, the isospin symmetry is restored in the $K$ projection.

\begin{figure}[tb]
\begin{center}
\includegraphics[width=4.5cm]{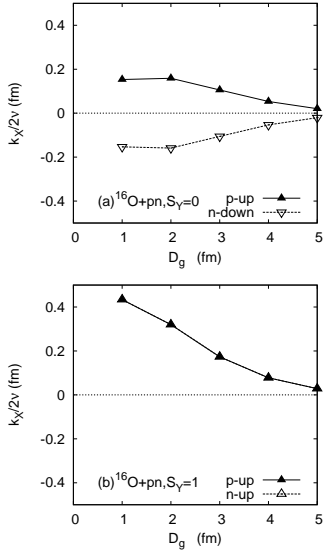} 	
\end{center}
\vspace{0.5cm}
  \caption{
The $D_g$ dependence of $k_{jX}$ obtained by the energy variation for 
the intrinsic wave function $\Phi_{^{16}{\rm O}+pn}(\bvec{R}_1,\bvec{R}_2)$ 
without the parity and angular momentum projections.  
Upper: $k_{jX}$ for the $\uparrow_Y\downarrow_Y$ $pn$ pair. 
Lower: $k_{jX}$ for the $\uparrow_Y\uparrow_Y$ $pn$ pair. 
In the lower panel, 
two lines for protons and neutrons overlap with each other
because $k_{1X}\approx k_{2X}$.  
The bh125 interaction with the spin-orbit force is used.
\label{fig:imx-nopro}}
\end{figure}

After the variation, the energy minimum state in the intrinsic system is obtained.
In the result of the $\uparrow_Y\downarrow_Y$ pair without the spin-orbit force, 
the optimized parameters for the energy minimum solution are 
$\bvec{R}_1\approx \bvec{R}_2\approx(0,0,D_{g})$ 
indicating that the $(0s)^2$ $pn$ pair is formed in the intrinsic system.
When the spin-orbit force is switched on, the optimized $\bvec{R}_j$ 
has the finite imaginary part as 
$\bvec{R}_1\approx(ik_X/2\nu,0,D_{g})$ and $\bvec{R}_2\approx(-ik_X/2\nu,0,D_{g})$
indicating the $pn$ pair localized around $(0,0,D_{g})$ consisting of 
$\uparrow_Y$ and $\downarrow_Y$ nucleons boosted in the opposite direction
with the finite momentum $k_X$.  The $k_X$ for the nucleon 
momentum in the $X$ direction is regarded as the 
order parameter as explained before, and  
the finite momentum $k_X$ means that the parity symmetry in the pair is 
broken because of the external spin-orbit field.
In Fig.~\ref{fig:imx-nopro}(a), we show the $D_{g}$ dependence of the nucleon 
momentum $k_{jX}$ in the $\uparrow_Y\downarrow_Y$ $pn$ pair.
$k_X=k_{1X}=-k_{2X}$ is the largest at the distance $D_{g}=2$ fm from the core 
and it gradually decreases as $D_{g}$ increases because the spin-orbit potential from the core 
gets weak with the increase of $D_{g}$.

For the $\uparrow_Y\uparrow_Y$ pair, we take off the constraint 
${\rm Im}[R_{gX}]=0$
and perform the energy variation with the constraints 
${\rm Re}[R_{1X}]=-{\rm Re}[R_{2X}]$, 
$R_{1Y}=-R_{2Y}$ and $R_{1Z}=R_{1Z}=D_{g}$ because 
two nucleons can be boosted in the same direction under the constraint
of ${\rm Im}[R_{gX}]=0$.
As expected from the role of the spin orbit field boosting nucleons at the surface, 
we obtain the minimum energy solution with 
$\bvec{R}_1\approx\bvec{R}_2\approx(i\kappa_X/2\nu,0,D_{g})$
indicating two nucleons boosted in the same direction which correspond to 
the $(0s)^2$ $pn$ pair rotating around the core 
with the finite momentum $K_{gX}=2\kappa_X$ of 
the c.m. motion of the pair.
The $D_{g}$ dependence of the optimized $k_{jX}$ for the $\uparrow_Y\uparrow_Y$ pair is 
shown in Fig.~\ref{fig:imx-nopro}(b). $\kappa_X=k_{1X}=k_{2X}$ decreases 
as $D_g$ increases.

\begin{figure}[tb]
\begin{center}
\includegraphics[width=6.5cm]{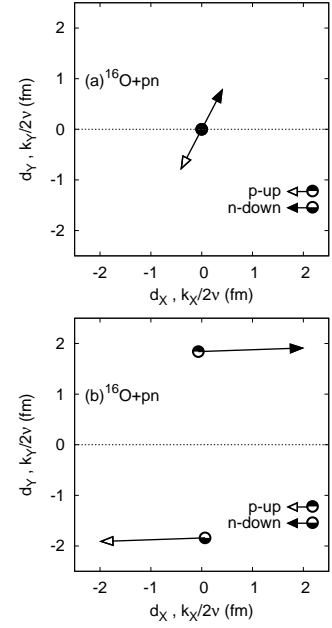} 	
\end{center}
\vspace{0.5cm}
  \caption{
$\bvec{R}_{1}$ and  $\bvec{R}_{2}$ in 
$\Phi_{^{16}{\rm O}+pn}(\bvec{R}_1,\bvec{R}_2)$ 
for the $\uparrow_Y\downarrow_Y$ $pn$ pair around $D_g=2$ fm
obtained by the VAP calculation for (a) the 
$T=1$ $J^\pi=0^+$ state and (b) the $T=0$ $J^\pi=1^+$ state.
The real parts $(d_{jX},d_{jY})$ are shown as the positions on 
the $XY$ plane, and the imaginary parts $(k_{jX}/2\nu,k_{jY}/2\nu)$ are illustrated by
the lengths of the arrows. 
The bh125 interaction with the spin-orbit force is used. 
\label{fig:sz0-spj}}
\end{figure}

We also perform the energy variation 
for the parity and total angular momentum projected states,
\begin{eqnarray}
&&\delta \frac{\langle \Psi|H|\Psi \rangle}{\langle \Psi|\Psi \rangle} =0,\\
&&\Psi=P^{J+}_{MK}\Phi_{^{16}{\rm O}+pn}(\bvec{R}_1,\bvec{R}_2).
\end{eqnarray}
Here $K=0$ and $K=1$ are chosen for $T=1$ and $T=0$ states, respectively. 
This is the variation after the projection (VAP). 
After the variation, the optimized parameters $\bvec{R}_1$ and $\bvec{R}_2$ of the energy minimum state
for the $J^\pi$ state are obtained. 
The VAP is performed for the $\uparrow_Y\downarrow_Y$ pair under the constraints 
$R_{1X}=-R_{2X}$, $R_{1Y}=-R_{2Y}$, and $R_{1Z}=R_{2Z}=D_{g}$.
$D_{g}$ is fixed to be 2 fm. 
The obtained results of $\bvec{R}_j=\bvec{d}_j+i\bvec{k}_j/2\nu$ 
for the $T=1$ $J^\pi=0^+$ state and those for the $T=0$ 
$J^\pi=1^+$ state are shown in Fig.~\ref{fig:sz0-spj}.
The position $\bvec{d}_j$ and the momentum $\bvec{k}_j$ are projected onto the $XY$ plane. 
Similarly to the variation without the projection, 
the $k_{jX}$ values obtained in the VAP for the $1^+$ and $0^+$ states 
are finite and they are opposite for two nucleons in the $\uparrow_Y\downarrow_Y$ pair 
so as to gain the spin-orbit potential.
It is consistent with the simple picture for the 
$\uparrow_Y\downarrow_Y$ pair where two nucleons are boosted along the $X$ direction in the opposite
direction. 
In addition, the $1^+$ and $0^+$ states obtained in the VAP have 
the finite $d_{jY}$ and $k_{jY}$ values, respectively. 
It indicates that the internal structure of the pair in the $J^\pi$ states
somewhat changes from the ideal $(0s)^2$ configuration because of core 
effects such as the central potential and also the antisymmetrization effects as well as the spin-orbit potential.
This means that the GCM calculation with only one generator coordinate of $R_X$ is not sufficient 
but that with two generator coordinates of $R_X$ and $R_Y$ is effective. 
Note that finite $d_{jY}$ and $k_{jY}$ are obtained only in the VAP but not in the 
variation without the projections in which the axial symmetry tends to be favored.


In spite of the finite $\bvec{R}_{jY}$ for the 
$\uparrow_Y\downarrow_Y$ pair, the momentum $k_X= k_{1X}=-k_{2X}$ in the $X$ direction 
is regarded as the order parameter for the symmetry breaking caused
by the spin-orbit field from the core.
To clarify the role of the spin-orbit force in the $pn$ pair, we perform further 
analysis of the $k_X$ dependence of the total energy of 
$\Phi_{^{16}{\rm O}+pn}(\bvec{R}_1,\bvec{R}_2)$ with fixed parameters 
$R_{jY}=0$ and $R_{jZ}=D_g$. 
In the following, we choose $D_g=2$ fm for the center position of the pair. 

\begin{figure}[tb]
\begin{center}
\includegraphics[width=7.5cm]{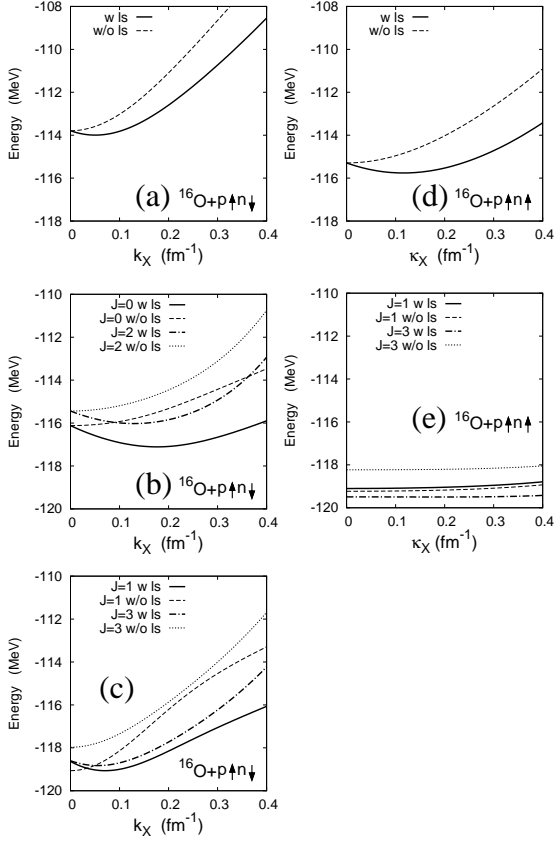} 	
\end{center}
\vspace{0.5cm}
  \caption{The energy 
of the $\Phi_{^{16}{\rm O}+pn}(\bvec{R}_1,\bvec{R}_2)$ wave function
for the $\uparrow_Y\downarrow_Y$ pair of two nucleons boosted in the opposite direction 
and that for $\uparrow_Y\uparrow_Y$ pair of two nucleons boosted in the same direction.
The parameters 
$\bvec{R}_1=(ik_X/2\nu,0,D_g)$ and $\bvec{R}_2=(-ik_X/2\nu,0,D_g)$ are used for the
$\uparrow_Y\downarrow_Y$ pair, and  
$\bvec{R}_1=(i\kappa_X/2\nu,0,D_g)$ and $\bvec{R}_2=(i\kappa_X/2\nu,0,D_g)$ are used for the
$\uparrow_Y\uparrow_Y$ pair.
$D_g$ is fixed to be 2 fm.
Left: the energy for the $\uparrow_Y\downarrow_Y$ $pn$ pair 
of (a) the intrinsic state without the $J^\pi$ projection, (b) the $J^\pi=0^+$ and $2^+$ projected states
with $T=1$, and (c) the $J^\pi=1^+$ and $3^+$ projected states with $T=0$.
Right: the energy for the $\uparrow_Y\uparrow_Y$ $pn$ pair 
of (d) the intrinsic state, and (e) the $J^\pi=1^+$ and $3^+$ 
projected states with $T=0$.
The bh125 interactions with and without the spin-orbit force are used.
\label{fig:d20}}
\end{figure}

We show in Fig.~\ref{fig:d20}(a)-(c) 
the energy of $\Phi_{^{16}{\rm O}+pn}(\bvec{R}_1,\bvec{R}_2)$ 
for the $\uparrow_Y\downarrow_Y$ $pn$ pair 
with the parameters $\bvec{R}_1=(ik_X/2\nu,0,D_g)$ ($D_g=2$ fm) 
and $\bvec{R}_2=(-ik_X/2\nu,0,D_g)$ as functions of $k_X$. It corresponds to the 
$pn$ pair localized around the position $(0,0,D_g)$ consisting of 
$\uparrow_Y$ and $\downarrow_Y$ nucleons 
boosted with the momentum $k_X$ in the opposite direction as shown 
in the upper panels of Fig.~\ref{fig:pn-pair}. The intrinsic wave function
$\Phi_{^{16}{\rm O}+pn}(\bvec{R}_1,\bvec{R}_2)$ 
for the $\uparrow_Y\downarrow_Y$ pair contains the $T=1$ and $T=0$ components
which are decomposed by the $K=0$ and $K=1$ projections. 
Energies for the intrinsic (no-projected) state are shown in 
Fig.~\ref{fig:d20}(a), and those for the $J^\pi$-projected states
are shown in Fig.~\ref{fig:d20}(b) and (c).
The $J^\pi$-projected energy is calculated by 
$P^{J\pi}_{MK}\Phi_{^{16}{\rm O}+pn}(\bvec{R}_1,\bvec{R}_2)$ with 
$K=0$ for the $J^\pi=0^+$ and $2^+$ states in the $T=1$ channel and
$K=1$ for the $J^\pi=1^+$ and $3^+$ states in the $T=0$ channel.

In the case without the spin-orbit force, 
the energy is minimum at $k_X=0$ corresponding to no symmetry breaking, i.e., 
no parity mixing in the pair for both the intrinsic and $J^\pi$ projected states.  
In the case with the spin-orbit force, 
the energy minimum of the intrinsic state
shifts to the finite $k_X$ region indicating that the symmetry is broken 
by the spin-orbit potential.
Further energy gain with the spin-orbit force is found 
in the $T=1$ $J^\pi=0^+$ and $2^+$ projected states.
In particular, the $T=1$ $J^\pi=0^+$ state largely gains 
the energy because of the spin-orbit force.
As shown in Fig.~\ref{fig:d20}(b) for the $T=1$ $J^\pi=0^+$ and $2^+$ states,
the spin-orbit potential energy is gained involving the parity mixing in the pair
in the finite $k_X$ region, but there is no spin-orbit potential contribution 
at the $k_X=0$ for the pure $(TS)=(10)$ state with which 
the expectation value of the spin-orbit potential vanishes.
The energy curve without the spin-orbit force is soft against the finite $k_X$ 
in the $T=1$ $J^\pi=0^+$ and $2^+$ states indicating that 
the internal energy loss in the $T=1$ pair with the finite $k_X$ is milder than that in the $T=0$ pair.
As a result, the finite $k_X$ state is favored and the large energy gain is obtained with 
the spin-orbit potential in
the $T=1$ $J^\pi=0^+$ and $2^+$ states.

On the other hand, in the $T=0$ $J^\pi=1^+$ state, 
the gain of the total energy because of the spin-orbit force is very small.
At $k_X=0$, the $T=0$ $J^\pi=1^+$ state is the pure $S=1$ state coupling with 
$L=0$ and $L=2$ waves of the c.m. motion of the pair and it 
feels no or slightly repulsive spin-orbit potential.
As $k_X$ increases, the energy without the spin-orbit force rapidly increases 
indicating the large internal energy loss in the $T=0$ pair in the $J^\pi=1^+$ state. 
This energy loss compensates the spin-orbit potential gain
resulting in almost no additional gain of the total energy in the spin-orbit potential.
It means that the finite $k_X$ states are unfavored in the $T=0$ $J^\pi=1^+$ state.
Also in the $T=0$ $J^\pi=3^+$ state, the finite $k_X$ states are not so favored.
However, differently from the $T=0$ $J^\pi=1^+$ state, 
the $T=0$ $J^\pi=3^+$ state at $k_X=0$ gains the spin-orbit force significantly 
because it is the $S=1$ state coupling mainly with 
the $L=2$ wave of the c.m. motion of the pair and feels the attractive spin-orbit 
potential.
Namely, for the $T=0$ states, 
the parity mixing in the pair does not contribute so much to the total 
energy gain in the spin-orbit
potential.

Let us discuss the $k_X$ dependence of the energy for the $T=0$ $\uparrow_Y\uparrow_Y$ pair
before and after the $J^\pi$ projection.
Note that the $J^\pi$ projected states 
for the $\uparrow_Y\uparrow_Y$ pair are almost equivalent to 
the $T=0$ $J^\pi$ projected states for the $\uparrow_Y\downarrow_Y$ pair with $k_X=0$ 
because both of them are pure $S=1$ states having 
the ideal $(0s)^2$ $pn$ pair.
We show in Fig.~\ref{fig:d20}(d) and (e) 
the energy of $\Phi_{^{16}{\rm O}+pn}(\bvec{R}_1,\bvec{R}_2)$ 
for the $\uparrow_Y\uparrow_Y$ pair 
with the parameters $\bvec{R}_1=\bvec{R}_2=(i\kappa_X/2\nu,0,D_g)$ ($D_g=2$ fm) 
meaning that two spin-up nucleons are 
boosted in the same direction and they form the $(0s)^2$ $T=0$ pair rotating around the core.
When the spin-orbit force is switched on, the energy minimum position shifts 
to the finite $\kappa_X$ region indicating that 
the c.m. motion of the $(TS)=(01)$ pair is boosted by the spin-orbit field.
Increase of $\kappa_X$ affects only the rotational mode of the c.m. motion of the pair 
keeping the internal structure unchanged. Since $\kappa_X$ changes only the proportion 
of the $J^\pi$ components contained in the intrinsic state, 
the $J^\pi$ projected energy does not depend 
on $\kappa_X$.  As shown in Fig.~\ref{fig:d20}(e), the $J^\pi=1^+$ and $3^+$
projected energies of for the $\uparrow_Y\uparrow_Y$ pair are almost independent from $\kappa_X$.  
Here we choose the angular momentum aligned to the intrinsic spin orientation in the projection as $J_Y=J$.


\begin{figure}[tb]
\begin{center}
\includegraphics[width=5.cm]{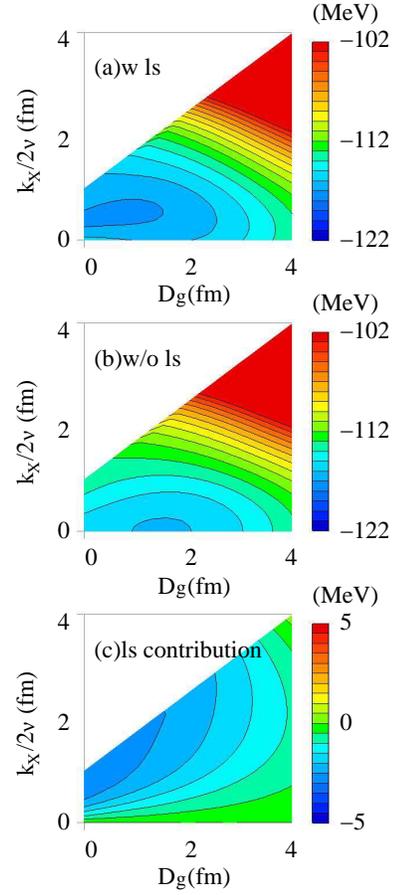} 	
\end{center}
\vspace{0.5cm}
  \caption{$k_X$ and $D_g$ dependence of the energy of 
the $\Phi_{^{16}{\rm O}+pn}$ wave function for the $\uparrow_Y\downarrow_Y$ 
$pn$ pair projected onto the $T=1$ $J^\pi=0^+$ state.
The  energy (a) with the spin-orbit force and (b) without 
the spin-orbit force, and (c) the contribution of the spin-orbit force evaluated by the energy difference between with and without 
the spin-orbit force.
\label{fig:j0-d-di}}
\end{figure}

\begin{figure}[tb]
\begin{center}
\includegraphics[width=5.0cm]{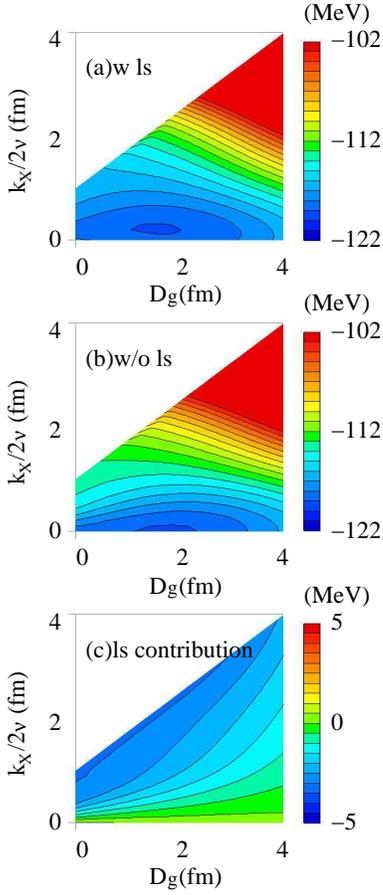} 	
\end{center}
\vspace{0.5cm}
  \caption{
  $k_X$ and $D_g$ dependence of the energy of 
the $\Phi_{^{16}{\rm O}+pn}$ wave function for the $\uparrow_Y\downarrow_Y$ 
$pn$ pair projected onto the $T=0$ $J^\pi=1^+$ state.
The  energy (a) with the spin-orbit force and (b) without 
the spin-orbit force, and (c) the contribution of the spin-orbit force evaluated by the energy difference between with and without 
the spin-orbit force.
\label{fig:j1-d-di}}
\end{figure}

Let us come back to the $\uparrow_Y\downarrow_Y$ $pn$ pair around the $^{16}$O core.
We discuss the $D_g$ 
dependence as well as the $k_X$ dependence of the energies of the $T=1$ $J^\pi=0^+$ and 
$T=0$ $J^\pi=1^+$ states projected from 
$\Phi_{^{16}{\rm O}+pn}(\bvec{R}_1,\bvec{R}_2)$ 
for the $\uparrow_Y\downarrow_Y$ $pn$ pair. We use the parametrization 
$\bvec{R}_1=(ik_X/2\nu,0,D_g)$ and $\bvec{R}_2=(-ik_X/2\nu,0,D_g)$.
The energies with and without the spin-orbit force are 
shown in Fig.~\ref{fig:j0-d-di} for the $T=1$ $J^\pi=0^+$ state
and in Fig.~\ref{fig:j1-d-di} for the $T=0$ $J^\pi=1^+$ state. 
The contribution of the spin-orbit force evaluated by the energy difference 
with and without the spin-orbit force is also shown. 
As shown in the bottom panels of Figs.~\ref{fig:j0-d-di} and \ref{fig:j1-d-di}, 
the spin-orbit force contribution is attractive in the finite $k_X$ region in both 
states, and the $k_X$ dependence of the attraction is not so different 
between the $0^+$ state and the $1^+$ state at least 
in $D_g\le 2$ fm region. The remarkable difference between 
the $0^+$ and $1^+$ states is found in the energy without the spin-orbit force.
In contrast to the large energy loss of the $1^+$ state in the finite $k_X$ region, 
the energy loss of the $0^+$ state is milder. As a result, in the total energy 
with the spin-orbit force, the minimum energy state for the 
$0^+$ state appears in the finite $k_X$ region and shows the significant reduction of the total energy
because of the spin-orbit force. 
Thus the parity symmetry is broken in the $T=1$ $pn$ pair by the spin-orbit field. In contrast,  
for the $1^+$ state, the energy minimum state in the total energy exists near the $k_X=0$ line 
suggesting smaller symmetry breaking in the $T=0$ $pn$ pair because of the large energy loss in the finite 
$k_X$ region.

As discussed in the previous section, the GCM calculation shows 
the smaller odd-parity component ${\cal P}_{\rm odd}$ of the $pn$ pair in 
the $T=0$ $J^\pi=1^+$ state than that in the $T=1$ $J^\pi=0^+$ state (see Table \ref{tab:18F-spin}).
It is consistent with the above analysis of the $k_X$ dependence of the energy 
with and without the spin-orbit force. 
The feature that the $T=0$ $J^\pi=1^+$ state is not favored in the spin-orbit 
potential originates in the internal energy loss caused 
by the parity mixing in the $T=0$ pair.
Although the $k_X$ dependence of the spin-orbit contribution is not 
so different between the $1^+$ and $0^+$ states,
the internal energy increases rapidly in 
the $T=0$ $J^\pi=1^+$ state than in the $T=1$ $J^\pi=0^+$ state 
as $k_X$ increases even in the case with
the equal $^3$E and $^1$E central forces. 
As explained before, for the $\uparrow_Y\downarrow_Y$ pair in the spin-orbit potential, 
the finite $k_X$ is favored to gain the spin-orbit potential 
energy roughly $-2\frac{\bar{U}_{ls}D_{g}}{2}k_{X}$ in the intrinsic frame. 
However, the finite $k_X$ inevitably causes the internal energy loss because of the 
mixing of the odd-parity component in the pair. 
The $T=1$ and $T=0$ components of the $\uparrow_Y\downarrow_Y$ pair are 
decomposed by the $K=0$ and $K=1$ projections, respectively.
For a given finite value of $k_X$, 
the internal energy loss of the pair is less 
in the $T=1$ pair than in the $T=0$ pair because the odd-parity mixing becomes about half of the intrinsic state in the $K=0$ projection for the $T=1$ component 
but it is not the case in the $K=1$ projection for the $T=0$ component 
as explained in appendix \ref{sec:app1}.
This means that, the finite $k_X$ state is unlikely in the $T=0$ pair
because of the larger internal energy loss than in the $T=1$ pair
though the $k_X$ dependence of the spin-orbit potential energy gain 
is almost equal in the $T=1$ and $T=0$ components.
As a result, in the energy minimum $T=0$ $J^\pi=1^+$ state with the spin-orbit potential, 
the parity mixing is suppressed and the total energy gain due to the spin-orbit potential 
is small.

\begin{figure}[tb]
\begin{center}
\includegraphics[width=5.0cm]{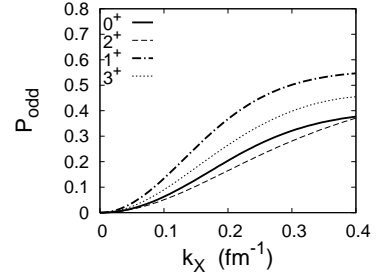} 	
\end{center}
\vspace{0.5cm}
  \caption{$k_X$ dependence of the odd-parity component ${\cal P}_{\rm odd}$
  in the $\Phi_{^{16}{\rm O}+pn}$ wave function for the $\uparrow_Y\downarrow_Y$ 
$pn$ pair projected onto the $T=0$ $J^\pi=1^+$ and $2^+$ states
and the $T=1$ $J^\pi=1^+$ and $3^+$ states. $D_g$ is fixed to be 2 fm.
\label{fig:d20-s-exp}}
\end{figure}

We compare the $k_X$ dependence of the odd-parity component  ${\cal P}_{\rm odd}$
in the $T=1$ and $T=0$ states projected from $\Phi_{^{16}{\rm O}+pn}$ wave function 
for the $\uparrow_Y\downarrow_Y$ 
pair in Fig.~\ref{fig:d20-s-exp}.
We use the parameterization $\bvec{R}_1=(ik_X/2\nu,0,D_g)$ 
and $\bvec{R}_2=(-ik_X/2\nu,0,D_g)$ with the fixed $D_g=2$ fm. 
It is found that, as $k_X$ increases, 
the odd-parity component in the $T=0$ $J^\pi=1^+$ state increases 
more rapidly than that in the $T=1$ $J^\pi=0^+$ state. 
In the small $k_X$ region, the odd-parity component in the $T=1$ $J^\pi=0^+$ state
is about half of that in the $T=0$ $J^\pi=1^+$ consistently with the reduction of the
odd-parity component in the $K$ projection described in appendix \ref{sec:app1}. 
The increase of the odd-parity component directly causes the internal 
energy loss of the pair. 
This is consistent with the arguments of Refs.\cite{Poves:1998br,Baroni:2009eh,Bertsch:2009xz,Sagawa:2012ta}
discussed from the mean field picture that the $T=0$ pairing 
is unfavored because of the
small overlap between the $jj$ coupling pair and the $LS$ coupling pair in the $T=0$ $J^\pi=1^+$ channel
than the $T=1$ $J^\pi=0^+$ channel.


\section{4-nucleon correlation at nuclear surface}\label{sec:4-nucleon}
 
\begin{figure}[tb]
\begin{center}
\includegraphics[width=6.5cm]{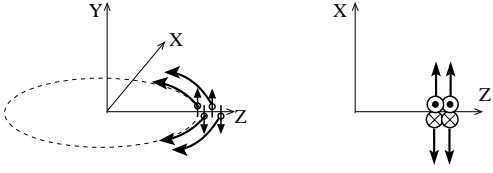} 	
\end{center}
\vspace{0.5cm}
  \caption{
Schematic figures for four nucleons in the spin-orbit potential
  at the nuclear surface.
\label{fig:4N}}
\end{figure}

 We discuss here the effect of the spin-orbit force on the 
$\alpha$ cluster breaking in analogy to the effect on the dinucleon pair.
As discussed before, the spin-orbit force changes the
internal structure of the $T=1$ $\uparrow_Y\downarrow_Y$ pair and 
the c.m. motion of the $T=0$ $\uparrow_Y\uparrow_Y$ pair in the intrinsic frame. 
The former contributes to the energy gain in the $T=1$ $J^\pi=0^+$ state
and the latter affects the energy gain of the 
$T=1$ $J^\pi=3^+$ and $5^+$ states. 
In contrast, the spin-orbit force gives minor contribution to the energy of 
the $T=0$ $J^\pi=1^+$ state.
For the $\alpha$ cluster around the $^{16}$O in the $T=0$ $J^\pi=0^+$ state,
a $\uparrow_Y\uparrow_Y$ $pn$ pair in a finite $L$ wave can couple with 
a $\downarrow_Y\downarrow_Y$ $pn$ pair in the $L$ wave in the opposite direction 
to form the $\alpha$ cluster in the total orbital angular momentum $L=0$ state
 (see Fig.~\ref{fig:4N}).
Therefore, the $\alpha$ cluster in the $L=0$ wave may gain the 
spin-orbit potential energy involving the 
cluster breaking from the $(0s)^4$ configuration.
It may be useful to discuss the effect of the 
spin-orbit force on the breaking in analogy to the effect on 
the $pn$ pair, although the $\alpha$ cluster breaking is expected to be 
suppressed because of the larger binding energy of the $\alpha$ cluster
than the $pn$ pair.

We perform analysis of four nucleons around the $^{16}$O 
by using  
the $^{16}$O+$4N$ wave function given in (\ref{eq:o16-4N})
in a similar way to the $pn$ pair. 
The adopted effective nuclear interaction is the bh125 interaction 
with and without the spin-orbit force.
We consider the $\alpha$ cluster localized around
the position $(0,0,D_g)$ on the $Z$-axis, 
\begin{equation} \label{eq:Rg-alpha}
\bvec{R}_g\equiv \frac{\bvec{R}_1+\bvec{R}_2+\bvec{R}_3+\bvec{R}_4}{4}=(0,0,D_g),
\end{equation}
with a real value $D_g$ for the distance from the $^{16}$O core. 
If we take $\bvec{R}_1=\bvec{R}_2=\bvec{R}_3=\bvec{R}_4=\bvec{R}_g$, 
four nucleons form the $(0s)^4$ $\alpha$ cluster.

Let us consider the $\alpha$ cluster breaking
in the spin-orbit field at the nuclear surface. 
We fix the $Z$-component ${R}_{jZ}=D_g$
and vary ${R}_{jX}$ and ${R}_{jY}$ to get the energy minimum state
under the constraints,
\begin{eqnarray}
&&\frac{R_{1X}+R_{2X}}{2}=\frac{R_{3X}+R_{4X}}{2}=0,\\
&&\frac{R_{1Y}+R_{2Y}}{2}=\frac{R_{3Y}+R_{4Y}}{2}=0.
\end{eqnarray}
This is equivalent to the constraints $R_{gX}=0$ and $R_{gY}=0$ 
without the dipole excitation. 
This model is regarded as
a special case of the d-constraint method in AMD \cite{Taniguchi:2004zz}.
After the energy variation, 
we obtain the optimum parameter set
$R_{jX}$ and $R_{jY}$ which minimizes the
energy of the wave function 
$\Phi_{^{16}{\rm O}+4N}(\bvec{R}_1,\bvec{R}_2,\bvec{R}_3,\bvec{R}_4)$ under 
the constraints. 

\begin{figure}[tb]
\begin{center}
\includegraphics[width=6.5cm]{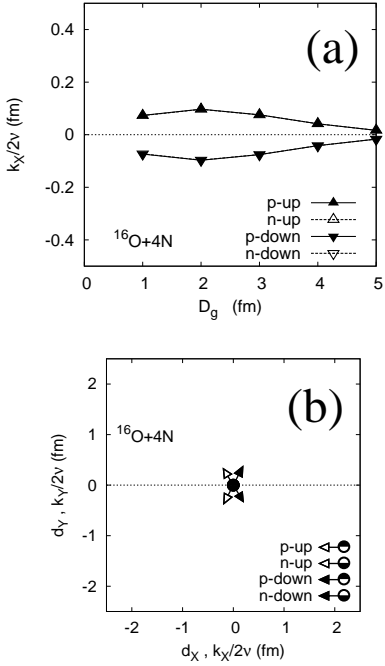} 	
\end{center}
\vspace{0.5cm}
  \caption{Optimized 
$\bvec{R}_{j}$ in 
$\Phi_{^{16}{\rm O}+4N}$ 
for the $\alpha$ cluster around the $^{16}$O core.
(a) $D_g$ dependence of $k_{jX}$ obtained by the energy variation 
without the projection. The $k_{jX}$ values 
for the $\uparrow_Y$ and $\downarrow_Y$ neutrons are consistent with the 
values for protons.  
(b) $\bvec{R}_{j}$
obtained by the VAP calculation for the 
$J^\pi=0^+$ state are projected onto the $XY$ plane. 
The real parts $(d_{jX},d_{jY})$ for four nucleons 
are shown as the position on the $XY$ plane, and they are located 
at the origin in the present result.
The imaginary parts $(k_{jX}/2\nu,k_{jY}/2\nu)$ for 
four nucleons are illustrated
by the lengths of the arrows. 
The bh125 interaction with the spin-orbit force is used.   
\label{fig:o16-4N}}
\end{figure}

We first perform the variation 
with respect to the intrinsic energy without 
the parity and the angular momentum projections.
In the result without the spin-orbit force, 
the optimum parameters in the minimum energy state are found to be  
\begin{equation}
\bvec{R}_{1}= \bvec{R}_{2}=\bvec{R}_{3}=\bvec{R}_{4}=
(0,0,D_g), 
\end{equation}
indicating that the ideal $(0s)^4$ $\alpha$ cluster is formed at the surface.
When the spin-orbit force is switched on, the optimum 
$\bvec{R}_j$ in the minimum energy state has
the finite imaginary part as 
$\bvec{R}_1=\bvec{R}_3\approx (ik_X/2\nu,0.D_g)$ and 
$\bvec{R}_2=\bvec{R}_4\approx (-ik_X/2\nu,0.D_g)$. 
It means that spin-up and -down nucleons in the $\alpha$ cluster 
are boosted in the opposite direction along the $X$-axis in the spin-orbit 
field from the $^{16}$O as shown in Fig.~\ref{fig:4N}.
The present result without the projections is consistent with the 
cluster breaking discussed in the AMD calculation of $^{28}$Si~\cite{KanadaEn'yo:2004cv}. 
It is also consistent with the simplified model for the $\alpha$-cluster breaking 
proposed by Itagaki {\it et al.} \cite{Itagaki:2005sy}.
The parameter $k_X$ in the present model relates to the 
parameter $\Lambda$ introduced in the simplified model as $\Lambda=k_X/2\nu D_g$, and 
it is regarded as the order parameter which indicates the cluster 
breaking at the nuclear surface. 
In Fig.~\ref{fig:o16-4N}(a), the $D_g$ dependence of the nucleon momentum
$k_{jX}$ obtained by the energy variation without the projections is shown. 
$k_X=k_{1X}=-k_{2X}=k_{3X}=-k_{4X}$ is largest at $D_g=2$ fm and it becomes small with the increase of 
the distance $D_g$ from the core. 

We also perform the energy variation for the $J^\pi=0^+$ state projected  
from the $\Phi_{^{16}{\rm O}+4N}$ wave function 
with $D_g=2$ fm. The parameters $\bvec{R}_j$ for 
nucleons in the $\alpha$ cluster obtained by the variation after projection 
(VAP) are shown in Fig.~\ref{fig:o16-4N}(b). 
In the VAP result, $k_{jY}$ is also finite as well as $k_{jX}$ indicating further
breaking of the $(0s)^4$ $\alpha$ cluster in addition to the breaking described by 
the parameter $k_{jX}$.
Qualitatively, the breaking of the $(0s)^4$ $\alpha$ cluster 
breaking around $^{16}$O core 
is similar to that of the $(0s)^2$ $pn$ pair in the $T=0$ $J^\pi=0^+$ state. 
However, quantitatively, 
the $\alpha$ cluster breaking is smaller than the $pn$ pair breaking
because the $\alpha$ cluster has the larger 
internal binding energy than the $pn$ pair and its breaking 
is unlikely.

\begin{figure}[tb]
\begin{center}
\includegraphics[width=6.5cm]{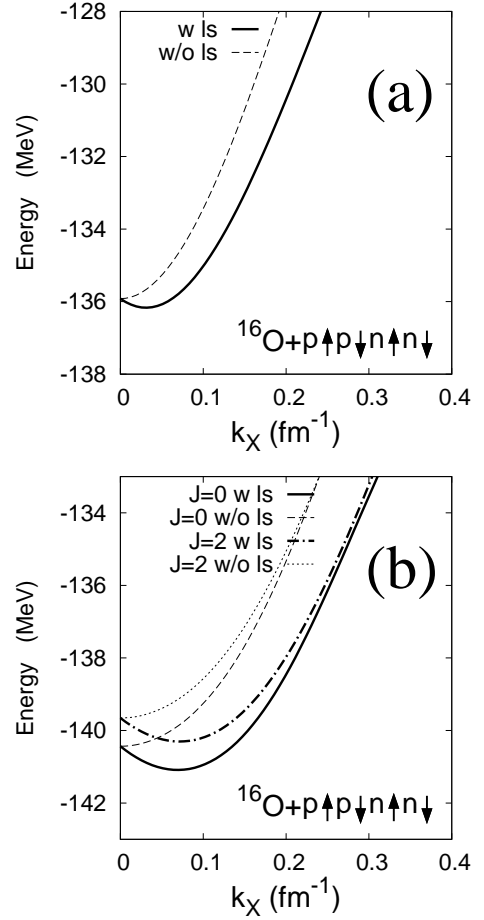} 	
\end{center}
\vspace{0.5cm}
  \caption{
$k_X$ dependence of the energy 
of the $\Phi_{^{16}{\rm O}+4N}$ wave function with  
$D_g=2$ fm.
(a) The energy of the intrinsic state, and (b) 
that of the $J^\pi=0^+$ and $2^+$ projected states.
The bh125 interactions with and without the spin-orbit force are used.
\label{fig:o16-4N-d20}}
\end{figure}

To see the energy gain of the spin-orbit potential in $\Phi_{^{16}{\rm O}+4N}$
with the cluster breaking, we analyze the $k_{X}$ dependence of the energy assuming 
${\rm Re}[R_{jX}]=0$ and $R_{jY}=0$ for simplicity. 
The calculated energies of the body-fixed intrinsic state and that of the $0^+$ state 
are shown in Fig.~\ref{fig:o16-4N-d20}. 
It is found that the energy minimum shifts to the finite $k_X$ region 
indicating that the $\alpha$ breaking occurs because of the spin-orbit field, 
in particular, in the $J^\pi=0$ projected states.
Compared with the $k_X$ dependence of the energy of the 
$\uparrow_Y\downarrow_Y$ $pn$ pair,  
the energy without the spin-orbit force is very steep with 
respect to the cluster breaking parameter $k_X$
because of the large internal energy loss of the $\alpha$ cluster.
As a result, the $\alpha$ cluster breaking in the $0^+$ state
is not as significant as the $\uparrow_Y\downarrow_Y$ $pn$ pair 
though the spin-orbit force gives some attractive effect 
in the finite $k_X$ region. 

It is interesting to consider the analogy of the effect of the  
spin-orbit force on 
the $\alpha$ cluster breaking
with that on the $pn$ pair discussed in the previous section.
The $\alpha$ cluster is composed of four nucleons, spin-up and -down 
protons and neutrons. 
In the spin-orbit field, the spin-up and -down nucleons are boosted in the opposite 
direction. It means that, in the body-fixed frame, 
the $\alpha$ cluster can be regarded as a composite of the 
$\uparrow_Y\uparrow_Y$ $pn$ pair and $\downarrow_Y\downarrow_Y$ $pn$ pair
which are boosted in the opposite direction (see Fig.~\ref{fig:4N}).
We also consider alternative interpretation of the $\alpha$ cluster as the 
composite of two $T=1$ pairs, $\uparrow_Y\downarrow_Y$ $pp$ and $nn$ pairs. 
Then, the behavior of the $T=1$ pairs in the 
$\alpha$ cluster is qualitatively consistent with the 
$\uparrow_Y\downarrow_Y$ $pn$ pair having the parity mixing in the intrinsic frame
discussed before. 

It should be commented that one should not directly link the $J^\pi=0^+$ $pp$ and 
$J^\pi=0^+$ $nn$ pairs
with the $\alpha$ cluster because 
the $\alpha$ cluster consists of correlating two pairs 
and it contains not only $J^\pi=0^+$ pairs but also finite $J$ pairs. 
In other words, $J^\pi=0^+$ $pp$ and $nn$ pairs have no spacial correlation 
between each other and they are different from the correlating four 
nucleons in the $\alpha$ cluster. 
We can connect the $4N$ correlation with the $2N$ correlation 
only in the intrinsic body-fixed frame before the $J$ projection  
or with the correlating pairs
having finite momenta.

%


\section{Summary}\label{sec:summary}

We investigated the structure of $^{18}$F with the microscopic wave function based
on the three-body $^{16}$O+$p$+$n$ model to discuss the behavior of the 
$pn$ pair around the $^{16}$O.
Particular attention was paid on the effect of the spin-orbit force on the 
$pn$ pair behavior.
 
In the GCM calculation, the $T=0$ energy spectra of 
$J^\pi=1^+$, $3^+$, and $5^+$ states and the $T=1$ spectra of 
$J^\pi=0^+$, $2^+$ states in $^{18}$F are described reasonably. 
The spin-orbit potential from the core plays an important role
in the energy spectra.
The spin-orbit potential energy gain is largest in the $T=0$ $J^\pi=5^+$ state
while it is small in the $T=0$ $J^\pi=1^+$ state. 
The significant energy gain of the spin-orbit potential is 
caused in the $T=1$ $J^\pi=0^+$states.

We discuss the effect of the spin-orbit force 
on the $T=0$ and $T=1$ $pn$ pair around the $^{16}$O based on the dinucleon
picture.
For the spin parallel $\uparrow_Y\uparrow_Y$ pair for the $T=0$ states, 
the spin-orbit potential boosts 
the c.m. motion of the pair in the rotational mode 
keeping the internal structure of the pair unchanged. It results in 
the larger energy gain in the higher spin $J$ states in the $T=0$ 
spectra.
For the spin antiparallel $\uparrow_Y\downarrow_Y$ pair in the body-fixed frame, 
the parity mixing in the pair occurs
because of the external spin-orbit field. In other words, 
the $pn$ pair gains the spin-orbit potential energy involving the 
odd-parity mixing, i.e., the symmetry breaking in the pair.
The spin-orbit potential energy gain with the odd-parity mixing 
is efficient in the $T=1$ pair in the $J^\pi=0^+$ state, but it is not 
so efficient in the $T=0$ pair in the $J^\pi=1^+$ state 
because of the large internal energy loss of the $T=0$ pair. 
The main origin of the smaller internal energy loss of the $T=1$ pair
than the $T=0$ pair is that 
the odd-parity component is reduced in the projection onto 
the $T=1$ $J^\pi=0^+$ eigen state but there is no reduction in the 
$T=0$ $J^\pi=1^+$ projection. 
Thus, the parity mixing is likely in the $T=1$ pair in the $J^\pi=0^+$
state but it is unlikely in the $T=0$ pair in the $J^\pi=1^+$ state. 
Since the $T=1$ $J^\pi=0^+$ pair is favored by the spin-orbit potential
while the $T=0$ $J^\pi=1^+$ pair is not favored, the $T=1$ $J^\pi=0^+$ state
comes down to the low energy region 
though the $T=0$ $J^\pi=1^+$ state is still the 
lowest state in $^{18}$F because of the 
stronger $^3$E nuclear force than the $^1$E force. 
  
The mechanism of the unfavorable $T=0$ pair in the spin-orbit potential 
is consistent with the discussions 
in Refs.~\cite{Poves:1998br,Baroni:2009eh,Bertsch:2009xz,Gezerlis:2011rh,Sagawa:2012ta,Bertsch:2009zd}.
One of the new standpoints in the present work is that 
we focus on the internal pair wave function and discuss its change 
involving the odd-parity mixing because of the spin-orbit force. 
In the present picture, the parity mixing in the pair is regarded
as the symmetry breaking in the pair by the symmetry variant external 
spin-orbit field. 

\section*{Acknowledgments} 
The authors would like to thank Dr.~Itagaki and Dr.~Tanimura for useful discussions.
The computational calculations of this work were performed by using the
supercomputers at YITP. This work was supported by 
JSPS KAKENHI Grant Numbers 22540275, 26400270.

\appendix

\section{Dinucleon cluster picture and parity mixing in the pair} \label{sec:app1}

We consider two nucleons around the core nucleus and describe  
how the parity symmetry in the pair is broken 
by the spin-orbit external field. 

We introduce a simplified potential model where 
all core effects are assumed to be renormalized 
in the one-body effective central and spin-orbit potentials while 
the recoil effect is omitted. 
In the simplified model, the Hamiltonian
is give as follows,
\begin{eqnarray}
&&H=t_1+t_2+U_c(r_1)+U_c(r_2)\nonumber\\
&&+U_{\rm ls}(r_1)\bvec{l}_1\cdot\bvec{s}_1+U_{\rm ls}(r_2)\bvec{l}_2
\cdot\bvec{s}_2\nonumber\\
&& +v_{NN}(r),
\end{eqnarray}
which describes interacting two nucleons in the external field from the core.
We first explain the mean-field picture and then describe 
the dinucleon picture based on this model for a pair around the core. 

In the mean field approximation, the Hamiltonian is rewritten as 
\begin{eqnarray}
&&H=h_1+h_2
+v_{NN}(r),\\
&& h_i=t_i+U_i,\\
&& U_i=U_c(r_i)+U_{\rm ls}(r_i)\bvec{l}_i\cdot\bvec{s}_i.
\end{eqnarray} 
The one-body parts of the Hamiltonian is considered to be the 
unperturbative Hamiltonian $H_0=h_1+h_2$ and 
the residual interaction $H'=v_{NN}$ is regarded as the 
perturbative Hamiltonian which causes the two-body correlation. 
In the leading term $H_0$, two nucleons behave as independent particles 
in the mean field $U_i$ containing the spin-orbit potential, 
and the correlated wave function is expressed by the linear combination of the
single particle configurations. It corresponds to the expression 
of the $jj$ coupling scheme.

When we respect the internal symmetry of the pair, the picture based on the 
$LS$ coupling scheme is useful rather than the $jj$ coupling scheme
because the Hamiltonian without the spin-orbit potential has the symmetry for the 
internal parity of the pair, which is explicitly broken by the spin-orbit potential.
For the $LS$ coupling picture, we consider another choice of the unperturbative 
Hamiltonian by regarding the spin-orbit potential as the
perturbative external field for two nucleons, 
\begin{eqnarray}
&&H=\tilde H_0+\tilde H',\\
&&\tilde H_0=t_1+t_2+U_c(r_1)+U_c(r_2)+v_{NN}(r),\\
&&\tilde H'=U_{\rm ls}(r_1)\bvec{l}_1\cdot\bvec{s}_1+U_{\rm ls}(r_2)\bvec{l}_2\cdot\bvec{s}_2.
\end{eqnarray} 
In $\tilde H_0$, two nucleons bound in the central potential 
form the $pn$ pair at the surface
because of the $S$-wave attraction of the nuclear force $v_{NN}$. The $(TS)=(01)$
and $(10)$ pairs are formed by the attractions in the $^3$E and 
$^1$E channels, respectively. 
Since the total intrinsic spin $S$ and the total orbital angular momentum $L$ are 
conserved, this picture is called the "$LS$ coupling" scheme. 
In the internal wave function of the pair, 
the parity transformation $\bvec{r}\rightarrow -\bvec{r}$ ($\bvec{r}\equiv \bvec{r}_1-\bvec{r}_2$)
is equivalent to the exchange $\bvec{r}_1\leftrightarrow \bvec{r}_2$,
and therefore, the internal parity of the pair 
is conserved in the unperturbative system 
because $\tilde {H}_0$
is invariant under the internal 
parity transformation, $\bvec{r}_1\leftrightarrow\bvec{r}_2$.

Based on the two-nucleon pair with the 
parity symmetry in the $LS$ coupling scheme,
we consider the spin-orbit potential as the perturbative external field that 
explicitly breaks the parity symmetry in the pair. 
The total Hamiltonian $H=\tilde H_0+\tilde H'$
is no longer invariant under the transformation
 $\bvec{r}_1\leftrightarrow \bvec{r}_2$.
As a result, the odd-parity component mixes in the dominant 
even-parity component in the internal pair wave function.
Because of the Fermi statistics, 
it means that the spin-singlet odd ($^1$O) component is mixed in the $^3$E component in the $T=0$ pair
and the $^3$O component is mixed in the $^1$E component in the $T=0$ pair.

To consider the breaking of the parity symmetry in the $pn$ pair, 
we discuss a $\uparrow_Y\downarrow_Y$ $pn$ pair 
localized around a certain position $(0,0,D_{g})$ on the $Z$ axis. 
Ignoring the degree of freedom along $Z$ axis
for simplicity, we look only into the two-dimensional (2D) problem on the $XY$ plane passing through
$(0,0,D_{g})$. For a simple explanation of the parity mixing 
in the pair due to the spin-orbit potential, 
we introduce a toy model
which mimics the 2D problem for the pair.
Namely, we assume the 2D H.O. potential around the origin of the coordinates $\bvec{\rho}=(X,Y)$, 
and add the perturbative field contributed by the spin-orbit potential.
\begin{eqnarray}
&&H_{\rm 2D}=H^{(0)}_{\rm 2D}+H'_{\rm 2D}, \\
&&H^{(0)}_{\rm 2D}=\frac{1}{2m} (p_{1X}^2+p_{1Y}^2)+\frac{m\omega^2}{2} \bvec{\rho}_1^2 \nonumber \\ 
&& +\frac{1}{2m} (p_{2X}^2+p_{2Y}^2)+\frac{m\omega^2}{2} \bvec{\rho}_2^2, \\ 
&& H'_{\rm 2D}=-\frac{\bar{V}_{\rm ls}}{\hbar}
(p_{1X} s_{1Y}-p_{1Y} s_{1X}) \nonumber\\
&& -\frac{\bar{V}_{\rm ls}}{\hbar} (p_{2X} s_{2Y}-p_{2Y} s_{2X}).
\end{eqnarray}
Here, the contributions from the nucleon-nucleon interaction as well as the central potential in $\tilde{H}_0$ are assumed to be renormalized in 
the H.O.-type mean potential $H^{(0)}_{\rm 2D}$. 
The spin-orbit potential contribution
is approximated by the averaged strength $-\bar{V}_{\rm ls}\approx U_{\rm ls}(D_g) D_g$  
assuming the coordinate $\bvec{r}_i$ in the spin-orbit potential to be constant 
$\bvec{r}_i \approx (0,0,D_g)$. $\bar{V}_{\rm ls}$ is a positive constant value. 
The H.O. assumption for the unperturbative potential is not essential but it can be
other potential form with 
the 2D-rotational symmetry (the axial symmetry in 3D).
The present H.O. assumption is used just for the convenience 
that the internal wave function and the c.m. wave function of the pair 
are separable in the lowest state of the H.O. potential. 

The lowest state of two nucleons, $\uparrow_Y p$ and $\uparrow_Y n$, 
for $H_0^{\rm 2D}$ 
is the $(0s)_{\rm 2D}^2$ configuration in 2D, 
\begin{eqnarray}
&&\Phi_0(1,2)=\frac{1}{2}{\cal A}\{ \phi_{0}(\bvec{\rho}_1) \chi_{p\uparrow_Y} 
\phi_{0}(\bvec{\rho}_2) \chi_{n\downarrow_Y} \},\\
&& \phi_0(\bvec{\rho}_i)=\phi^{\rm 2D}_{0s}(b;\bvec{\rho}_i),\\ 
&& b= \sqrt{\frac{\hbar}{m\omega}},
\end{eqnarray}
where $\phi^{\rm 2D}_{0s}(b;\bvec{\rho})$ is the 
function for the $0s$ state of the 2D H.O. with the size parameter 
$b$ 
\begin{equation}
\phi^{\rm 2D}_{0s}(b;\bvec{\rho})
\equiv\left( \frac{1}{\pi b^2} \right)^{1/2} e^{-\frac{\rho^2}{2b^2}}.
\end{equation}
The spatial part of the wave function of the $(0s)^2_{\rm 2D}$ state is expressed 
by a product of the c.m. wave function $\phi^{2D}_{g,0}$ and the internal wave function $\phi^{2D}_{{\rm in},0}$, 
\begin{eqnarray}
&& \phi_{0}(\bvec{\rho}_1) \phi_{0}(\bvec{\rho}_2) = \phi_{g,0}(\bvec{\rho}_g) \phi_{{\rm in},0} (\bvec{\rho}) ,\\
&& \phi^{2D}_{g,0}=\phi^{\rm 2D}_{0s}(\frac{b}{\sqrt{2}};\bvec{\rho}_g),\\
&& \phi^{2D}_{{\rm in},0}=\phi^{\rm 2D}_{0s}(\sqrt{2}b;\bvec{\rho}),
\end{eqnarray}
with the c.m. and relative coordinates, $\bvec{r}_{g}\equiv (\bvec{r}_1+\bvec{r}_2)/2$ and $\bvec{r}\equiv \bvec{r}_1-\bvec{r}_2$ of the pair. 
Needless to say, the internal wave function  $\phi^{2D}_{{\rm in},0}$
of the pair contains only the even-parity component.

Because of the spin-momentum coupling term $H'_{\rm 2D}$ originating in the 
spin-orbit potential, the parity mixing occurs in the  $\uparrow_Y\downarrow_Y$ pair. 
The total Hamiltonian for the $\uparrow_Y\downarrow_Y$ pair can be written as 
\begin{eqnarray}
&& H^{\rm 2D}=\frac{1}{2m} \left\{ 
(p_{1X}- \hbar k_X)^2 + p_{1Y}^2 \right \}
+ \frac{m\omega^2}{2} \bvec{\rho}_1^2 \nonumber\\
&& +\frac{1}{2m} \left\{ 
(p_{2X}+ \hbar k_X)^2 + p_{2Y}^2 \right \}
+ \frac{m\omega^2}{2} \bvec{\rho}_2^2 \nonumber\\
&&+ C ,\\
&&k_X\equiv \frac{m}{2\hbar^2}\bar{V}_{\rm ls} ,\\
&&C=-\frac{m}{4\hbar^2}\bar{V}^2_{\rm ls}.
\end{eqnarray} 
The energy shift $C$ of the lowest state is proportional to $\bar{V}^2_{\rm ls}$
instead of $\bar{V}_{\rm ls}$ because the leading order of the energy perturbation vanishes,
$\langle\Phi_0|H'_{2D}|\Phi_0\rangle=0$. 
Single particle wave functions for the spin-up ($\uparrow_Y$) and -down 
($\downarrow_Y$) nucleons in the lowest state 
are those shifted in the momentum space in the opposite direction along the $X$ axis, 
\begin{eqnarray}
&&\phi(\bvec{\rho}_1)=\left( \frac{1}{\pi b^2} \right)^{1/2} e^{-\frac{\rho_1^2}{2b^2}} e^{ik_X X_1},\\
&&\phi(\bvec{\rho}_2)=\left( \frac{1}{\pi b^2} \right)^{1/2} e^{-\frac{\rho_2^2}{2b^2}} e^{-ik_X X_2}.
\end{eqnarray}
In other words, two nucleons are boosted with the momentum $k_X$ in the opposite direction by the 
spin-momentum coupling field. As a result 
the odd-parity mixing occurs in the internal 
wave function of the pair as
\begin{eqnarray}
&&\phi(\bvec{\rho}_1)\phi(\bvec{\rho}_2)=\phi^{2D}_{g,0}(\bvec{\rho}_{g}) \phi^{2D}_{\rm in}(\bvec{\rho}),\\ 
&&\phi^{2D}_{\rm in}(\rho) = \left( \frac{1}{2\pi b^2} \right)^{1/2} e^{-\frac{\bvec{\rho}^2}{4b^2}} e^{ik_X X}\nonumber\\
&&=\phi^{2D}_{0s}(\sqrt{2}b:\bvec{\rho}) +  ik_Xb\phi^{2D}_{0p_X}(\sqrt{2}b;\bvec{\rho}) 
+ \mathcal{O}(k^2_X).\nonumber\\
\end{eqnarray}
Here $\phi^{2D}_{0p_X}$ indicates the $0p_X$ state in the 2D H.O.
In the limit of the small perturbation, the odd-parity component is 
$k_X^2/4b^2 = m^2 \bar{V}^2_{\rm ls}/16\hbar^4 b^2$ which is proportional to the square of the 
strength of the spin-orbit field. 

Clearly shown in this schematic model, 
the parity symmetry in the pair breaks because of the spin-orbit field. 
The order parameter 
$k_X$ relating to the odd-parity component is determined by
the competition between the spin-orbit potential energy gain and 
the energy cost to break the symmetry. 
The energy cost comes from the 
energy loss to excite the $0p$ state from the lowest $0s$ state in the pair,
and it is proportional to the $0p$ mixing (parity mixing) component.
In the following, we show that the ratio of the $0p$ component to the 
dominant $0s$ component in the pair changes in the $J_Z=K$ projection 
which is equivalent to the $T$ projection. 
Namely, the $0p$ component is quenched in the $T=1$ projection resulting in the
smaller internal energy loss in the $T=1$ pair.  
Similar discussions was done in Ref.~\cite{Bertsch:2009zd}.


The $\uparrow_Y\downarrow_Y$ $pn$ pair contains $T=0$ and $T=1$ components as 
\begin{eqnarray}
\chi_{p\uparrow_Y} \chi_{n\downarrow_Y}= \frac{{\cal X}_{01}}{2} +\frac{{\cal X}_{10}}{2} 
+ \frac{{\cal X}_{00}}{2} +\frac{{\cal X}_{11}}{2}, 
\end{eqnarray}
where  ${\cal X}_{TS}$ is the isospin-spin $TS$ state of two nucleons. 
The parity mixed $pn$ pair can be decomposed into $T=0$ and $T=1$ eigen states as, 
\begin{eqnarray}
&&\Phi(1,2)=\frac{1}{2}{\cal A}\{ \phi(\bvec{\rho}_1) \chi_{p\uparrow_Y} 
\phi(\bvec{\rho}_2) \chi_{n\downarrow_Y} \}\nonumber\\
&&= \phi^{\rm 2D}_{g,0}(\bvec{r}_g)\cdot \nonumber\\
&&\left\{ \phi^{2D}_{0s}(\sqrt{2}b;\bvec{\rho}) \frac{{\cal X}_{01}}{2} 
+ ik_Xb \phi^{2D}_{0p_X}(\sqrt{2}b;\bvec{\rho}) \frac{{\cal X}_{00}}{2} \right.\nonumber\\
&&
\left. +\phi^{2D}_{0s}(\sqrt{2}b;\bvec{\rho}) \frac{{\cal X}_{10}}{2} 
+ ik_Xb \phi^{2D}_{0p_X}(\sqrt{2}b;\bvec{\rho}) \frac{{\cal X}_{11}}{2}
 \right\}\nonumber\\
&&+\mathcal{O}(k^2_X).
\end{eqnarray}
In the 2D system in the intrinsic frame, 
there is no difference in the ratio of the even-parity and odd-parity components
between $T=0$ and $T=1$ states. However, since the 2D Hamiltonian has the rotational symmetry around the $Z$ axis (the axial symmetry), and as a result of 
the symmetry restoration, the $J_Z=K$ eigen state projected 
from the intrinsic state should be
considered.
The $K=0$ and $K=\pm 1$ projected pairs correspond to the $T=1$ and $T=0$ pairs, 
respectively, 
and they are naively expected in the $T=1$ $J^\pi=0^+$ 
and $T=1$ $J^\pi=1^+$ states. 
Choosing the $Z$ axis as the quantization axis, the orbital angular momentum $0p_X$ state
can be rewritten in terms of $|ll_Z\rangle_{l}$ as
\begin{eqnarray}
|0p_X\rangle = \frac{1}{\sqrt{2}}\left(|1-1\rangle_l - |1+1 \rangle_l \right),
\end{eqnarray}  
and also the $S=1$ component of the $S_Y=0$ state in ${\cal X}_{T1}$ 
can be rewritten in terms of $|SS_Z\rangle_{s}$ as 
\begin{eqnarray}
|{\cal X}_{T1}\rangle = \frac{i}{\sqrt{2}}\left(|1-1\rangle_s + |1+1 \rangle_s \right) |T\rangle.
\end{eqnarray}  
For the $T=1$ pair, the odd-parity term 
$\phi^{2D}_{0p_X}(2b) {\cal X}_{11}$ 
can be decomposed by $|1-1\rangle_l |1-1\rangle_s $,
$|1+1\rangle_l |1+1\rangle_s $,
$|1-1\rangle_l |1+1\rangle_s $, and
$|1+1\rangle_l |1-1\rangle_s $ components.
It means that the odd-parity term of the $T=1$ component 
contains the $K=0$ and $K=2$ components in equal weight, 
while the even-parity term $\phi^{2D}_{0s}(2b){\cal X}_{10}$ contains only the $K=0$
component. As a result, the odd-parity component is suppressed 
in the $K=0$ projected state, and the ratio of the odd-parity to the even-parity
components in the 
$T=1$ pair is reduced to be half of the intrinsic state.
On the other hand, for the $T=0$ pair, there is no reduction of the odd-parity 
component in the $K=\pm 1$ projection, and the ratio in the $T=0$ pair 
is the same as that in the intrinsic state, 
In other words, the even-parity component 
is relatively enhanced in the $K=0$ projection for the $T=1$ state, 
while such the enhancement of the even-parity component 
does not occur in the $K=\pm 1$ projection for 
the $T=0$ state.

\section{Relation between $pn$ cluster wave function and shell-model wave function}\label{sec:app2}

In this work, we use the three-body cluster model of $^{16}$O$+p+n$ where 
with the form of Gaussian wave packets for two nucleons. 
We here show that this wave function becomes 
a $sd$-shell configuration of the H.O. shell model in a certain limit. 

In a basis cluster wave function $\Phi_{^{16}{\rm O}+pn}(\bvec{R}_1,\bvec{R}_2)$ 
of Eq.~\ref{eq:o-pn}, the spatial part of 
the single-particle wave function for the $i$th valence nucleon
is described with a Gaussian wave packet of Eq.~\ref{eq:single-phi},
\begin{equation}
\phi(\boldsymbol{R}_j; \boldsymbol{r}_i) = \left( \frac{2\nu}{\pi} \right)^{3/4}
e^{-\nu (\boldsymbol{r}_i - \boldsymbol{R}_j)^2}.
\end{equation}
Using the expansion $e^{-t^2+2xt} = \sum_{n=0}^{\infty} H_n(x) t^n/n!$
with the Hermite polynomial $H_n(x)$, 
this Gaussian wave packet can be rewritten by an expansion of 
the H.O. shell-model single particle wave function $(n_x,n_y,n_z)_{\rm ho}$ 
for the width parameter $b=1/\sqrt{2\nu}$ as 
\begin{align}
&\phi(\boldsymbol{R}_j; \boldsymbol{r}_i) = \sum_{n_x,n_y,n_z} {\cal N}_{n_xn_yn_z}(\boldsymbol{R}_{j})(n_x,n_y,n_z)_{\rm ho} \\
&  {\cal N}_{n_xn_yn_z}(\boldsymbol{R}_{j}) =
e^{- \frac{\nu}{2} \boldsymbol{R}_{j}^2} 
\prod_{\sigma = x,y,z}
\left( \sqrt{\frac{\nu}{2}} R_{j \sigma} \right)^{n_\sigma}
\frac{2^{n_\sigma/4}}{(n_\sigma !)^{3/4}}.
\end{align}
Note that the coefficient ${\cal N}_{n_xn_yn_z}(\boldsymbol{R}_{j})$ is the order of 
$|\boldsymbol{R}_{j}|^N$ ($N\equiv n_x+n_y+n_z$) in the small 
$|\boldsymbol{R}_j|$ limit.

Since $N\le 1$ orbits are already occupied by nucleons in the $^{16}$O core, 
only $N\ge 2$ orbits are allowed for valence nucleons in the 
the total wave function $\Phi_{^{16}{\rm O}+pn}(\bvec{R}_1,\bvec{R}_2)$. 
It means that $N< 2$ shell-model configurations vanish in the total wave function 
after the antisymmetrization and $N\le 2$ configurations remains.
Defining the orthogonal component of $\phi(\boldsymbol{R}_j; \boldsymbol{r}_i)$
to $N\le 1$ orbits  as 
\begin{equation}
\tilde \phi(\boldsymbol{R}_j; \boldsymbol{r}_i)=
 \sum_{N \ge 2} {\cal N}_{n_xn_yn_z}(\boldsymbol{R}_{j})(n_x,n_y,n_z)_{\rm ho}, 
\end{equation}
we can rewrite the $^{16}$O$+p+n$ wave function as 
\begin{eqnarray}
\Phi_{^{16}{\rm O}+pn}(\bvec{R}_1,\bvec{R}_2)&=&  {\cal A} 
\left\{ 
\Phi_{^{16}{\rm O}}\tilde \psi_{p\sigma}(\bvec{R}_1) \tilde \psi_{n\sigma'}(\bvec{R}_2) \right\},\\
\tilde \psi_{\tau\sigma}(\bvec{R}_j;\bvec{r}_i) 
&=&  \tilde \phi(\bvec{R}_j;\bvec{r}_i) \chi_{\tau\sigma}.
\end{eqnarray}
In the limit of $|\boldsymbol{R}_j| \rightarrow 0$, 
the lowest order $\tilde \phi(\boldsymbol{R}_j; \boldsymbol{r}_i)$ term becomes dominant in 
$\tilde \phi(\boldsymbol{R}_j; \boldsymbol{r}_i)$ as 
\begin{eqnarray}
\tilde \phi(\boldsymbol{R}_j; \boldsymbol{r}_i)&= 
 \sum_{N = 2} {\cal N}_{n_xn_yn_z}(\boldsymbol{R}_{j})(n_x,n_y,n_z)_{\rm ho}\\
&+ {\cal O}(|\boldsymbol{R}_j|^3).
\end{eqnarray}
This means that the 
$^{16}$O$+p+n$ cluster model wave function becomes 
a $(sd)^2$ configuration of the H.O. shell-model
wave function in the small $|\boldsymbol{R}_j|$ limit.

\end{document}